\documentclass{aa}

\usepackage{txfonts,textcomp}
\usepackage{graphicx}   
\usepackage{siunitx}            
\usepackage{enumitem} 
\usepackage[shortcuts,acronyms]{glossaries-extra}               
\usepackage{orcidlink}

\newabbr{GC}{GC}{globular cluster}
\newabbr{SC}{SC}{stellar cluster}
\newabbr[longplural = supernovae, shortplural = SNe]{SN}{SN}{supernova}
\newabbr{MF}{MF}{stellar mass function}
\newabbr{IMF}{IMF}{initial mass function}
\newabbr{SF}{SF}{star formation}
\newabbr{SFE}{SFE}{star formation efficiency}
\newabbr{SFR}{SFR}{star formation rate}
\newabbr{MW}{MW}{Milky Way}
\newabbr{AGB}{AGB}{asymptotic giant branch}
\newabbr{SMBH}{SMBH}{super-massive black hole}
\newabbr{BH}{BH}{black hole}
\newabbr{NS}{NS}{neutron star}
\newabbr{T1}{T1}{tail 1}
\newabbr{T2}{T2}{tail 2}
\newabbr{MLR}{MLR}{mass-luminosity relation}
\newabbr{MLE}{MLE}{maximum likelihood estimator}

\newcommand{\Kpc}{\, \mathrm{kpc}}
\newcommand{\Msun}{\, \mathrm{M}_{\odot}}
\newcommand{\Myr}{\, \mathrm{Myr}}
\newcommand{\Kms}{\, \mathrm{km} \, \, \mathrm{s}^{-1}}
\newcommand{\nbdvi}{\textsc{nbody6} }

\begin{document}

\title{The masses of open star clusters and their tidal tails and the stellar initial mass function}

\author{
Henriette Wirth\orcidlink{0000-0003-1258-3162}$^{1}$\thanks{E-mail: wirth@sirrah.troja.mff.cuni.cz (HW)},
František Dinnbier\orcidlink{0000-0001-5532-4211}$^{1}$,
Pavel Kroupa\orcidlink{0000-0002-7301-3377}$^{1,2}$
and Ladislav Šubr\orcidlink{0000-0003-1924-8834}$^1$}

\institute{$^{1}$Charles University, Faculty of Mathematics and Physics, Astronomical Institute, V Hole\v{s}ovi\v{c}kách 2, Praha, CZ-18000, Czech Republic\\
$^{2}$Helmholtz Institut für Strahlen und Kernphysik, Universität Bonn, Nussallee 1416, 53115 Bonn, Germany}
             
\date{Accepted XXX. Received YYY; in original form ZZZ}

 
\abstract
   {Unresolved binaries have a strong influence on the observed parameters of \acp{SC}.}
   {We quantify this influence and compute the resulting mass underestimates and \ac{MF}.}
   {N-body simulations of realistic \acp{SC} were used to investigate the evolution of the binary population in a \ac{SC} and its tidal tails.
Together with an empirically gauged stellar mass-luminosity relation, the results were then used to determine how the presence of binaries changes the photometric mass and \ac{MF} of the \ac{SC} and its tails as deduced from observations.}
   {\Ac{T1}, which is the tidal tail caused by gas expulsion, contains a larger fraction of binaries than both the \ac{SC} and \ac{T2}, which forms after gas expulsion.
Additionally, \ac{T1} has a larger velocity dispersion.
Using the luminosity of an unresolved binary, an observer would underestimate its mass.
This bias sensitively depends on the companion masses due to the structure of the stellar mass-luminosity relation.
Combining the effect of all binaries in the simulation, the total photometric mass of the \ac{SC} is underestimated by 15\%.
Dark objects (black holes and neutron stars) increase the difference between the real and observed mass of the \ac{SC} further.
For both the \ac{SC} and the tails, the observed power-law index of the \ac{MF} between a stellar mass of 0.3 and 0.7 $M_\odot$ is smaller by up to 0.2 than the real one, the real \ac{IMF} being steeper by this amount.
This difference is larger for stars with a larger velocity dispersion or binary fraction.}
   {Since the stars formed in \acp{SC} are the progenitors of the Galactic field stars, this work suggests that the binary fractions of different populations of stars in the Galactic disc will differ as a function of the velocity dispersion.
   However, the direction of this correlation is currently unclear, and a complete population synthesis will be needed to investigate this effect.
   Variations in the binary fractions of different clusters can lead to perceived variations of the deduced stellar \acp{MF}.}


\titlerunning{Biases in the stellar mass function}
\authorrunning{Henriette Wirth, František Dinnbier, Pavel Kroupa,
and Ladislav Šubr}

\maketitle




\section{Introduction}
\label{sec_intro}

Stellar clusters (SCs)  are the building blocks of our and other galaxies \citep{1995MNRAS.277.1491K,2002MNRAS.336.1188K,2003ARA&A..41...57L,2005ESASP.576..629K,2010ARA&A..48..431P}.
Therefore, understanding their masses and the mass functions (MFs) within them is of the utmost importance for comprehending the Universe as we see it today.
One factor that strongly influences the measurements of these properties are unresolved binaries \citep{1991MNRAS.251..293K,2008A&A...480..103K,2010MNRAS.402.1750G,2018arXiv180610605K,2019ApJ...874..127B,2020ApJ...896..152R}.

While the binary fraction in the Galaxy is $\approx 0.5$ \citep[][and references therein]{1991A&A...248..485D,2010ApJS..190....1R,2011MNRAS.417.1702M}, \acp{SC} are likely born with up to 100 \% of their stars in binaries \citep{1994ARA&A..32..465M,1995MNRAS.277.1491K,1995MNRAS.277.1507K}.
The difference between the observed distribution functions of binary systems in the Galactic field in comparison to that observed in star-forming regions allowed \citet{1995MNRAS.277.1491K,1995MNRAS.277.1507K} to deduce the properties of the initial distribution functions of young binary-rich populations.
There is still some debate around whether this is also true for very low-mass binaries, many of which might dissolve before reaching the main sequence or even not start out as binaries \citep{2015ApJ...800...72T,2023ASPC..534..275O}.
Due to the dense environment of the clusters, the initially high binary fractions in \acp{SC} are reduced quickly \citep{1995MNRAS.277.1491K,1995MNRAS.277.1507K,2005MNRAS.358..572I,2005A&A...439..565G,2007ApJ...670..747K}.
The stars lost during \ac{SC} evolution form the field stars observed in the Galaxy today \citep{2002MNRAS.330..707K,2011MNRAS.417.1702M,2022MNRAS.510..413D,2023arXiv230105166F}.

\cite{2020A&A...640A..84D} categorised these stars into tail 1 and tail 2 (\ac{T1} and \ac{T2}) classes.
 \ac{T1} stars are lost from the birth-embedded cluster due to the decrease in the gravitational potential during gas expulsion.
The \ac{T2} stars, on the other hand, evaporate slowly from the \ac{SC} after this initial phase.
Both regimes overlap in phase space and so are not observed as two separated regions.
They are also contaminated by high-velocity stars ejected due to few-body interactions and supernova (SN) kicks.
These tails are expected to inherit binaries from the \ac{SC}, and if the binaries are unresolved, this has implications for the observations of the tails and the \ac{SC}.

One of the most common ways to determine the mass of a star is to use the measured luminosity.
Many studies have determined the \ac{MLR} for main-sequence stars in different mass ranges \citep[see e.g.][and references within]{1923BAN.....2...15H,1923PASP...35..189R,1993MNRAS.262..545K,1993AJ....106..773H,2015AJ....149..131E}.
However, if a binary is unresolved, only the combined luminosity of its members can be measured.
An unassuming observer would take this combined luminosity and use it to compute the mass of the assumed single star.
This mass, however, does not equal the sum of the masses of the binary components.
Therefore, an observer would arrive at a wrong mass for the binary and the entire \ac{SC}.

Alternatively, the mass of a virialised SC can be deduced dynamically because its dynamical mass is proportional to the square of the line-of-sight velocity dispersion \citep[e.g.][]{1976ApJ...204...73I,1987degc.book.....S}.
However, for this deduction, a binary fraction of 0 is assumed.
While this mass determination might still work if centre-of-mass velocities for binaries are used, many studies have shown that the velocity dispersions for systems with large fractions of unresolved binaries are overestimated \citep[see e.g.][]{2008A&A...480..103K,2010MNRAS.402.1750G,2019ApJ...874..127B,2020ApJ...896..152R,2021ApJ...908...60B}.
Therefore, binaries will cause an overestimate of the \ac{SC} mass if it is measured dynamically.
However, long-duration observations over multiple epochs can mitigate the effect \citep{2012A&A...539A...5C, 2024arXiv240808116R}.
\citet{2014A&A...562A..20C} suggest that two epochs are sufficient for this in many \acp{SC}; however, they also point out that many binaries have periods of thousands of years, making it practically impossible to detect them.

Similarly, the presence of binaries influences the deduced \ac{MF} for low-mass stars \citep{1991MNRAS.251..293K}.
Having an unresolved binary means that, when determining the \ac{MF}, two lower-mass stars are missing and one slightly more massive star is added \citep{2018arXiv180610605K}.
This makes the \ac{MF} appear more top-heavy.
\cite{2023arXiv230107029L} investigated this effect using mock data, and it was further investigated using N-body simulations in this work.

The purpose of this work is, therefore, to determine how the binary fraction affects our understanding of the mass and \ac{MF} in \acp{SC} and their tails.
For the first time, this is done using realistic N-body simulations of \acp{SC} to determine the development of the binary population.
Under the assumption that all binaries are unresolved, possible conclusions an observer would draw from these data are computed and compared to the results obtained under the assumption that all binaries are resolved.
This work is structured as follows:
Section \ref{sec_methods} explains the methods used for this work.
The results are discussed in Sect. \ref{sec_res}, and Sect. \ref{sec_concl} concludes the paper.

\section{Methods}
\label{sec_methods}

\subsection{Description of the simulations}

\begin{table}
    \caption{Description of the basic model parameters.}
    \label{tab_modelParam}
    \begin{center}
    \begin{tabular}{lcccc}
        \hline
        & {M1} & {M2} & {M3} & {M4}\\
        & {(main model)}\\\hline
        Number of models & 1 & 5 & 10 & 16\\
        Initial mass [$M_\odot$] & 6400 & 3200 & 1600 & 800\\
        $r_h [\mathrm{pc}]$ & 0.31 & 0.28 & 0.26 & 0.24\\
        $t_{\mathrm{gas}}$ [Myr] & 0.6 & 0.6 & 0.6 & 0.6\\
        $\overline{t_{\mathrm{ee}}}$ [Myr] & 81.6 & 82.4 & 90.0 & 94.9 \\
        $t_{\rm rlx}$ [Myr] & 58 & 40 & 28 & 19 \\\hline
    \end{tabular}
    \end{center}
    \tablefoot{For each set of models the number of models is given as well as the initial mass, half-mass radius, $r_h$, gas expulsion time, $t_{\rm gas}$, average threshold time, $\overline{t_{\mathrm{ee}}}$ and half-mass relaxation time, $t_{\rm rlx}$, calculated from Eq. (7.108) of \citet{Binney2008}.}
\end{table}

\begin{figure}
    \includegraphics[scale=1]{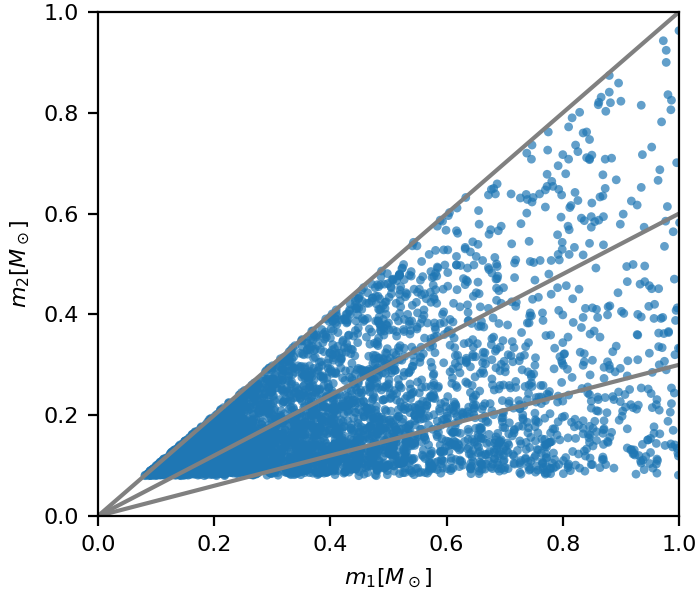}
    \caption{Mass of the primary, $m_1$, versus the mass of the secondary, $m_2$.
    The grey lines are (from top to bottom) for the cases $q = \frac{m_2}{m_1} = 1$, $q = 0.6,$ and $q = 0.3$.
    The points show stars drawn randomly from the canonical IMF.}
    \label{fig_q}
\end{figure}

The SC models initially follow non-mass-segregated Plummer density profiles and the  \citet{2012A&A...543A...8M} relation between the initial cluster mass in stars ($M_{\rm ecl}$) and the half-mass radius ($r_{\rm h}$).
This relation was deduced from the density at birth of open clusters required to account for the observed binary populations in open clusters.
It is important to emphasise that the \acp{SC} are non-mass-segregated, because the stars in the outer regions of the \acp{SC} are preferentially lost during gas expulsion and through evaporation.
On the other hand, dynamical ejection is more important in the \ac{SC} centre, as the high density of stars leads to a large number of encounters.
Since evaporation and gas expulsion are the main ways stars are lost, a mass-segregated \ac{SC} would lead to the tidal tails becoming bottom-heavy, while the \ac{SC} would become bottom-light.

Table \ref{tab_modelParam} lists the initial parameters of the main model, M1, used in this work and a number of lower-mass models, initialised with different random seeds, used for additional analysis.
Initially, the clusters are embedded in their natal gas, which is approximated by an analytical Plummer model of the same half-mass radius as the stellar population. 
The total gaseous mass is $2 M_{\rm ecl}$ in each model; this assumes a star formation efficiency of $1/3$ \citep[see e.g.][]{2016AJ....151....5M}.

Up to the age of $0.6 \Myr,$ the gaseous component is constant in time; after that, its density decreases exponentially on a timescale of $r_{\rm h}/(10 \Kms)$, which approximates feedback from young massive stars. 
More details of the gas expulsion model can be found in \citet[see also \citealt{Lada1984} and \citealt{Goodwin1997} for earlier analytic models of gas expulsion and \citealt{2015ApJ...814L..14C} and \citealt{2020MNRAS.499..748D} for hydrodynamical simulations]{2001MNRAS.321..699K}.
The clusters move on circular orbits within the Galactic disc at galactocentric radius $R_{\rm g} = 8 \Kpc$; the Galactic potential is represented by the \citet{Allen1991} model.

Initial positions and velocities of stars correspond to virial equilibrium with their primordial gas; they were generated using the \citet{Aarseth1974b} method.
Individual stellar masses were sampled from the two-part-power-law canonical initial mass function \citep[\ac{IMF};][]{2001MNRAS.322..231K, 2013pss5.book..115K}:
\begin{align}
    \xi(m) = \begin{cases}
                 k_1 m^{-1.3} & m < 0.5 ~M_\odot, \\
                 k_2 m^{-2.3} & m \geq 0.5 ~M_\odot.
             \end{cases}
\end{align}
Here $k_1$ and $k_2$ are constants chosen such that the \ac{IMF} is continuous and the integration over the \ac{IMF} yields the total mass of the \ac{SC}.
The stars have masses $m > 0.08 M_\odot$, brown dwarfs are neglected as they constitute a negligible mass contribution.

All stars are paired to binaries. 
To match observational quantities (e.g. \citealt{1991A&A...248..485D,Sana2012,Kobulnicky2014}), the distributions for mass ratio, orbital period, and eccentricity are separated into a set used for lower-mass (primary mass $m_1 < 3 \Msun$) and one for massive ($m_1 \geq 3 \Msun$) binaries \citep{1995MNRAS.277.1491K,1995MNRAS.277.1507K,2017MNRAS.471.2812B}.
We note that the boundary between high- and low-mass stars is at $5 \Msun$ in \citet{2017MNRAS.471.2812B}, while in the current work it is $3 \Msun$.
This has a negligible effect on the circularisation and secondary masses of stars in this range.
To compute the initial stellar properties, first the pre-main-sequence stars are generated following \citep{1995MNRAS.277.1507K} with the updates in \citet{2017MNRAS.471.2812B}:

The masses of the primary, $m_1$, and the secondary, $m_2$, are drawn randomly from the \citet{2001MNRAS.322..231K} \ac{IMF}.
For the high-mass stars ($m_1 \geq 3 \Msun$), stars are ordered before being paired: the most massive star is paired with the second most massive one, and so on.
The low-mass stars ($m_1 < 3 \Msun$) are paired randomly.
The fraction of orbits with an eccentricity between $e$ and $\mathrm{d}e$, $\mathrm{d}N = f_e \mathrm{d}e$, is described by Eq. 4b of \citet{1995MNRAS.277.1491K}:
\begin{align}
    f_e = 2e,
\end{align}
and the fraction of periods between $\log_{10} P$ and $\log_{10} P + \mathrm{d} \log_{10} P$, $dN = f_P \mathrm{d} \log_{10} P$ is described by Eq. 8 of \citet{1995MNRAS.277.1507K}:
\begin{align}
    f_p = 2.5 \frac{\log_{10} \left( P/\mathrm{days} \right) - 1 }{45 + \left[ \log_{10} \left( P/\mathrm{days} \right) - 1 \right]^2}.
\end{align}

For the massive stars, the pre-main-sequence phase is assumed to be too short to change the properties of the binaries significantly.
However, for the low-mass stars, the properties are changed due to pre-main-sequence eigenevolution as computed by \citet{1995MNRAS.277.1507K} and summarised by \citet{2017MNRAS.471.2812B}:
\begin{align}
    \rho &= \left( \frac{28 R_\odot}{R_p} \right)^\frac{3}{4},\\
    \ln (e_\mathrm{ini}) &= - \rho + \ln( e_\mathrm{bir} )\\
    q_\mathrm{ini} &= \begin{cases} q_\mathrm{bir} + \rho \left( 1 - q_\mathrm{bir} \right), & \rho \leq 1,\\
    1, & \rho > 1, \end{cases}\\
    m_{1, \mathrm{ini}} &= m_{1, \mathrm{bir}},\\
    m_{2, \mathrm{ini}} &= q_\mathrm{ini} m_{1, \mathrm{bir}},\\
    P_\mathrm{ini} &= P_\mathrm{bir} \sqrt{ \frac{ m_{1,\mathrm{bir}} + m_{2,\mathrm{bir}} }{ m_{1,\mathrm{ini}} + m_{2,\mathrm{ini}} }} \left( \frac{1 - e_\mathrm{bir}}{1 - e_\mathrm{ini}} \right)^\frac{3}{2}.
\end{align}
Here the subscript `ini' denotes that a variable is at the end of the eigenevolution during the pre-main-sequence time, while the subscript `bir', for birth value, is used for variables at the beginning of the pre-main-sequence phase. $R_p$ is the pericentre distance of the binary, $q \leq 1$ the fraction between secondary mass and primary mass, $m_1$ and $m_2$ are the primary and secondary mass, respectively, and $P$ the period of the binary.
We note that the secondary masses are slightly increased through pre-main-sequence eigenevolution, flattening the \ac{IMF} negligibly.

In this work we define the binary fraction as $f_{\rm bin} = \frac{N_{\rm bin}}{N_{\rm bin} + N_{\rm sing}}$. The multiple fraction is defined as
\begin{align}
    f_{\rm mul} = \frac{N_{\rm mul}}{N_{\rm mul} + N_{\rm sing}}.
    \label{eq_fmul}
\end{align}
Here $N_{\rm sing}$, $N_{\rm bin}$, and $N_{\rm mul}$ are the number of single stars, binary systems, and multiple systems, respectively.

The initial distribution of the masses of the two binary components for lower-mass binaries is shown in Fig. \ref{fig_q}.
It is important to note that the binaries are first generated from a list of stellar masses drawn from the \ac{IMF} and then assembled into a \ac{SC}.
Due to the dense environment inside the \ac{SC} this can lead to an overlap of binary pairs or to multiple pairs of binaries being extremely close to each other.
Therefore, not all of the stars originally set into the cluster as binaries turn out to be an actual binary pair as determined from their orbital energy.
Consequently, the actual binary fraction is only around 90\% in the initial timestep, $f_{\rm bin} = 0.9$.

The clusters in all models are evolved over a period of $\approx 300 ~\mathrm{Myr}$ by the code \nbdvi \citep{1999PASP..111.1333A,2003gnbs.book.....A}, which couples regularising techniques \citep{Kustaanheimo1965,Aarseth1974a,Mikkola1990} with a fourth-order Hermite predictor-corrector scheme \citep{Makino1991,Makino1992}.
The code contains synthetic evolutionary tracks for both single star and binary star evolution \citep{Tout1996,Hurley2000,Hurley2002}.
The velocities for the {SN} kicks are drawn from a Gaussian distribution with $\sigma = 40 ~\mathrm{km /s}$.

To distinguish between \ac{T1}, which is caused by gas expulsion, and \ac{T2}, caused by energy-equipartition-driven evaporation, Eq. 29 from \cite{2020A&A...640A..84D} was used to compute the threshold time, $t_\mathrm{ee}$.
The threshold time before which escaping stars are counted towards \ac{T1} is ${t_\mathrm{ee, main} = 81.6 ~\mathrm{Myr}}$ for the main model.
For the lower-mass models, we list the averages in Table \ref{tab_modelParam}.
We note that $t_\mathrm{ee}$ slightly varies between individual models.
For example for M2 $t_\mathrm{ee}$ ranges from ${78.0 ~\mathrm{Myr}}$ to ${86.8 ~\mathrm{Myr}}$.

\subsection{The computation of the observed masses}

\begin{figure}
    \includegraphics[scale=1]{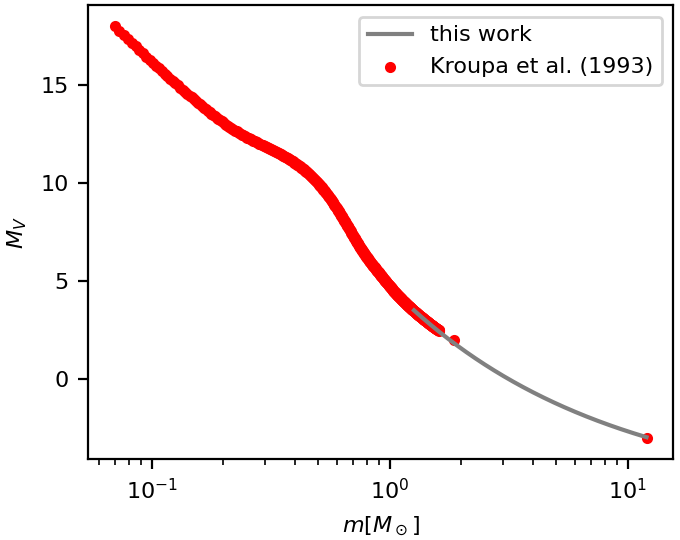}
    \caption{ MLR used in this work. The absolute V-band magnitudes of stars, $M_V$, are shown in relation to their stellar mass, $m$.
    The red dots show the data points from \citet{1993MNRAS.262..545K}, while the grey line shows the fit used for $m > 1.5 M_\odot$.}
    \label{fig_MassBrFunc}
\end{figure}

It is assumed that the observer is unable to resolve any of the binaries.
Since the stars spend most of their time on the main sequence the \ac{MLR} from \citet[][in their Appendix A]{1993MNRAS.262..545K} can be used to get the luminosity, $L(m)$, for each star \citep[see also][]{2002AJ....124.2721R}.
The main advantage of this \ac{MLR} over others \citep[e.g.][]{1923BAN.....2...15H,1923PASP...35..189R,2015AJ....149..131E} is that it extends down to very low-mass stars ($0.07~M_\odot$) and physically correctly accounts for the non-linearity near $0.35~M_\odot$.
The luminosity of the members of multiple systems is then added and converted back to a mass that represents the mass an observer would assume for the unresolved system:
\begin{align}
    m_\mathrm{obs} = L^{-1}\left( L(m_1) + L(m_2) \right),
\end{align}
with the masses of the individual stars, $m_1$ and $m_2$, and the apparent mass an observer that cannot resolve the binary would compute from the total luminosity, $m_\mathrm{obs}$.

Figure \ref{fig_MassBrFunc} shows the used \ac{MLR}.
The grey line follows the function $M_V = (10.8 \pm 0.1) \left( m/M_{\odot} \right)^{-0.491 \pm 0.006} + (- 6.1\pm0.1)$.
This function was fitted using the data points for $m > 0.96 M_\odot$ and is used to compute the luminosities for all stars with $m > 1.5 M_\odot$.
For all other masses the values of the closest red data point are used.
The magnitudes, $M_V$, are transformed to luminosities, $L$, using the following equation:
\begin{align}
        \label{eq_MagLumi}
    \frac{L(m)}{L_\odot} = 10^{-0.4 ( M_V( m ) - M_{V,\odot} )},
\end{align}
where $M_{V,\odot} = 4.83$ is the absolute V-band magnitude of the Sun.

It is important for our work to use an observationally derived \ac{MLR}, rather than a theoretical one, since the observationally derived \ac{MLR} correctly displays the KTG peak \citep{1990MNRAS.244...76K} of the stellar luminosity function at the transition from partially convective to fully convective stars \citep{2002Sci...295...82K}.
If not accounted for correctly, this leads to spurious results on the stellar \ac{MF} \citep{1990MNRAS.244...76K}.

In the following sections the results for both the main and the lower-mass models are shown.
The results for the lower-mass models are either averaged within their respective group (M2, M3, or M4; i.e. in the cases where quantities are shown over time) or the quantities of the lower-mass models are shown together (i.e. when the quantities of the individual binaries are shown).

\section{Results and discussion}
\label{sec_res}

\begin{table*}
    \begin{center}
        \caption{Results of the models discussed in the present work.        }
        \label{tab_results}
        \begin{tabular}{l|SSSScc}
            Model Name & {$\frac{M_{\mathrm{SC,obs}}}{M_{\mathrm{SC,tot}}} (t = 300 ~\mathrm{Myr})$} & {$\frac{M_{\mathrm{SC,dyn}}}{M_{\mathrm{SC,tot}}} (t = 300 ~\mathrm{Myr})$} & {$\sigma_{z, \mathrm{T1}} ~[\mathrm{km/s}]$} & {$\sigma_{z, \mathrm{T2}} ~[\mathrm{km/s}]$} & {$b$} & {$b_\mathrm{MLE}$} \\\hline
            M1 (main model) & 0.83 & 1.86 & 3.24 & 0.58 & $0.8 \pm 0.3$ & $0.9 \pm 0.2$ \\
            M2 & 0.82 & 2.81 & 2.81 & 0.54 & $1.5 \pm 0.3$ & $1.4 \pm 0.3$ \\
            M3 & 0.81 & 4.05 & 2.11 & 0.43 & $1.7 \pm 0.4$ & $1.2 \pm 0.4$ \\
            M4 & 0.78 & 17.52 & 1.70 & 0.27 & $1.8 \pm 0.4$ & $1.5 \pm 0.5$
        \end{tabular}
        \tablefoot{The table lists the model name, the fraction of the total model mass an observer would deduce when determining the luminous mass of the SC at the end of the simulation ($\frac{M_{\mathrm{SC,obs}}}{M_{\mathrm{SC,tot}}}$), the fraction of the total model mass an observer would deduce when determining the luminous mass of the SC at the end of the simulation ($\frac{M_{\mathrm{SC,dyn}}}{M_{\mathrm{SC,tot}}}$), and the velocity dispersion in the z-direction averaged over the last 100 Myr of the simulation for T1 and T2 ($\sigma_{z, \mathrm{T1}}$, and $\sigma_{z, \mathrm{T2}}$ respectively).
        Additionally, the factor b from Eq. \ref{eq_linfit} is listed for both the fitting method and the MLE.
        All values are averages of the results of all models within one group (M1, M2, M3, or M4).}
    \end{center}
\end{table*}

\subsection{The apparent stellar masses of unresolved multiples}
\label{sec_obsSysM}

\begin{figure}
    \includegraphics[scale=1]{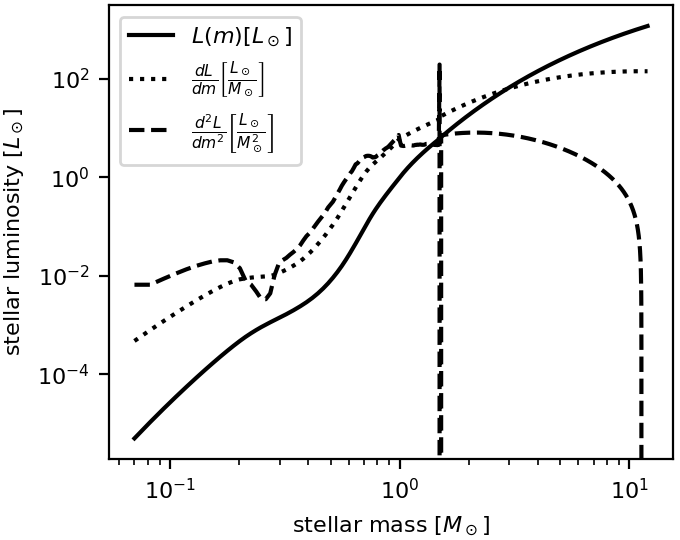}
    \caption{Luminosity of a star (Eq. \ref{eq_MagLumi} and Fig. \ref{fig_MassBrFunc}) and its first and second derivative with respect to the stellar mass as a function of the stellar mass.}
    \label{fig_MassFluxFunc}
\end{figure}

For the purpose of this discussion, all bright stars are assumed to be on the main sequence, that is to say, the short phases between when they are on the main sequence and when they become a \ac{BH}, neutron star (NS), or white dwarf are ignored.
Since the luminosity is added instead of the masses the relationship of the luminosity and the mass determines whether or not the mass of a system is over- or underestimated.

In Fig. \ref{fig_MassFluxFunc}, the $L(m)$ used in this work and its first and second derivative are shown.
We note that the anomaly in the second derivative at $1.5 ~M_\odot$ stems from the transition from using the next data point from the \ac{MLR} to using the fitted function.
This leads to a small jump in the derivative of $L(m)$, which is further enhanced through the second derivative.
Since the second derivative is positive, $L(m)$ must be a convex function in the given interval.

\begin{figure*}
    \includegraphics[scale=1]{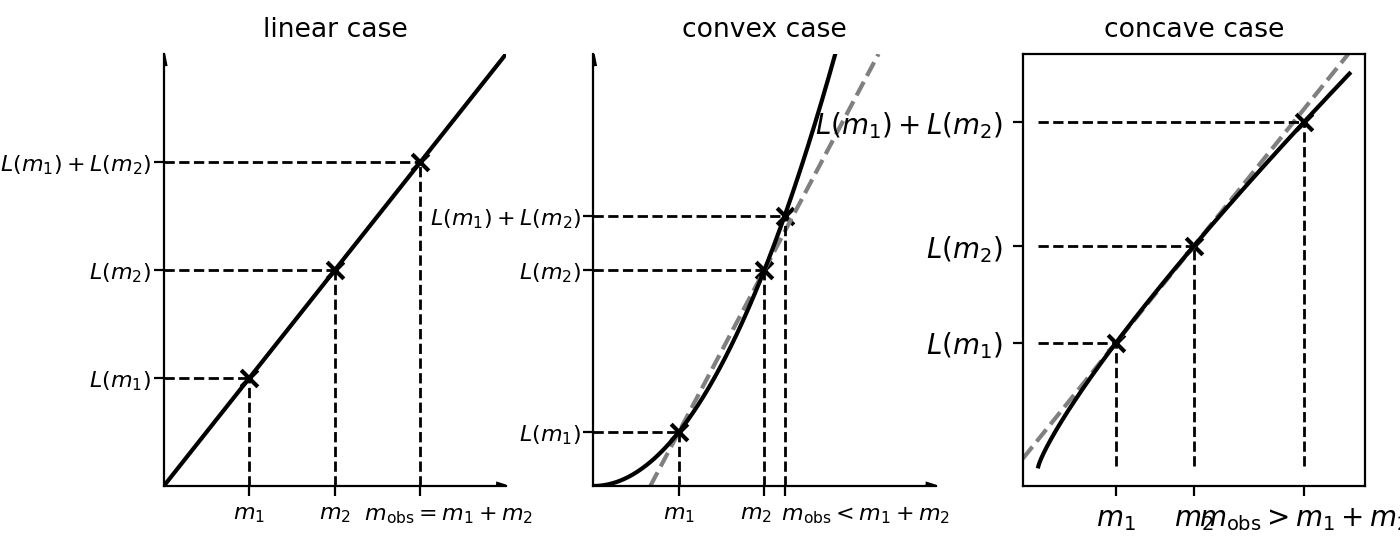}
    \caption{Examples on how the non-linearity of $L(m)$ can lead to an underestimate or an overestimate of the total mass of a binary star system.}
    \label{fig_Functions}
\end{figure*}

To understand, how this affects our estimate of the total mass of a binary, the following is considered:
Assume two stars with masses $m_1$ and $m_2$.
If the function is convex in an interval $[0, m_1 + m_2]$ (i.e. every point lies below its chord), where $L(0) = 0,$ it immediately follows that
\begin{align}
    \frac{L(m_1 + m_2)}{ m_1 + m_2 } &> \frac{L(m_1)}{m_1},\\
    \frac{L(m_1 + m_2)}{ m_1 + m_2 } &> \frac{L(m_2)}{m_2}.
\end{align}
Multiplying with the denominator on the right side results in\begin{align}
    L(m_1 + m_2) \frac{m_1}{m_1 + m_2} &> L(m_1),\\
    L(m_1 + m_2) \frac{m_2}{m_1 + m_2} &> L(m_2).
\end{align}
Adding both equations together leads to
\begin{align}
    L(m_1 + m_2) \frac{m_1 + m_2}{ m_1 + m_2} &> L(m_1) + L(m_2),\\
    L(m_1 + m_2) &> L(m_1) + L(m_2).
\end{align}
Thus, the luminosities of the individual stars added up is always going to be smaller than the total luminosity of a single star with mass $m_1 + m_2$.
This means that in this case the mass of an unresolved binary is always going to be underestimated.
Analogous it could be shown that for a concave function with $L(0) = 0$ the mass of an unresolved binary is always going to be overestimated.
An example of this can be seen in Fig. \ref{fig_Functions}.
It should be noted here that the $L(m)$ used in the present work is only known down to $0.08 M_\odot$.
This can affect whether the mass of the binaries is over- or underestimated especially for the low-mass stars.

\begin{figure*}
    \includegraphics[scale=1]{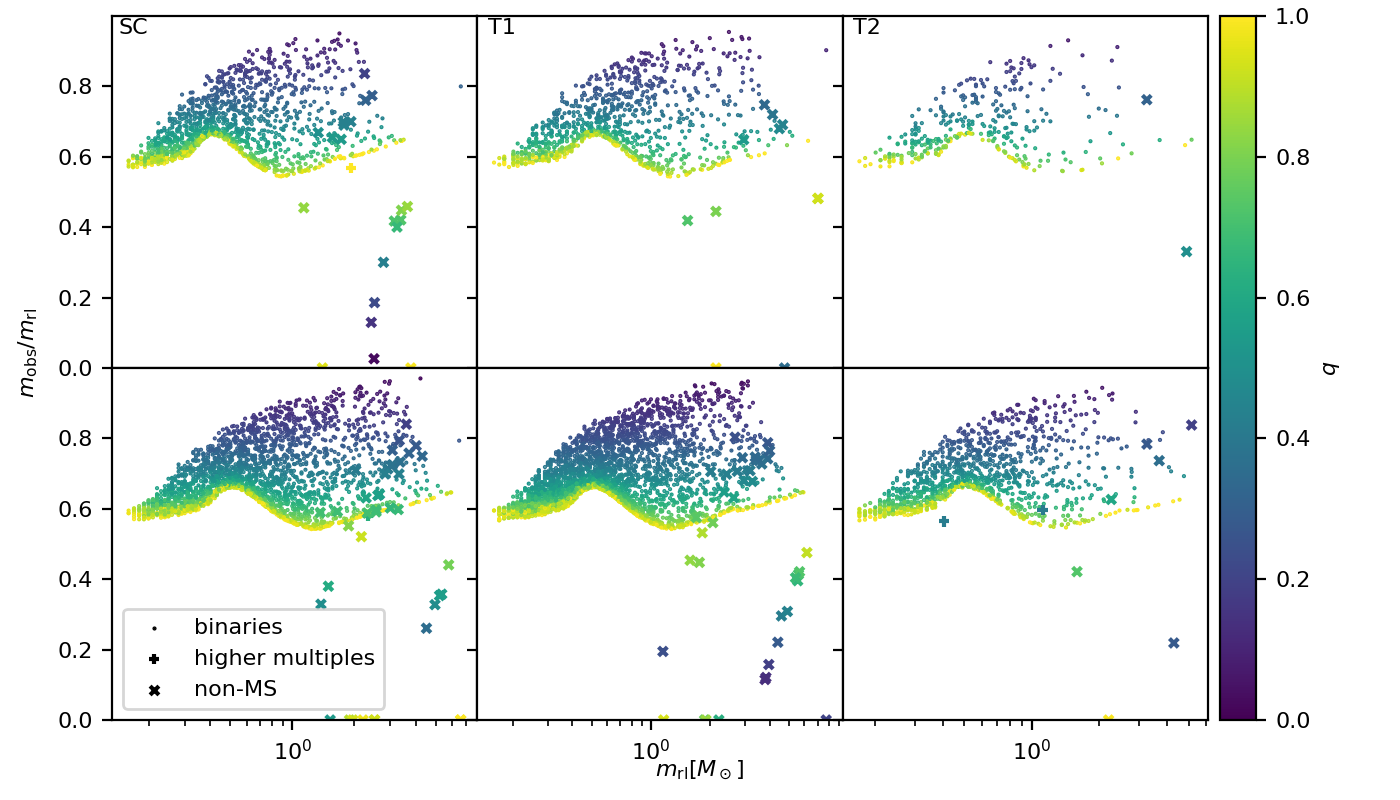}
    \caption{Masses of multiple star systems in the SC (left), T1 (centre), and T2 (right) of the main model (top) and the five M2 models (bottom) as seen by an observer over the real total mass of the systems at the end of the simulation ($T = 300 ~\mathrm{Myr}$).
    Binaries composed of only main-sequence stars are shown as dots, while multiple systems containing non-main-sequence stars are displayed as crosses.
    The colour denotes the mass ratio of the system.}
    \label{fig_BinMasses}
\end{figure*}

Figure \ref{fig_BinMasses} shows by how much an observer would underestimate the total mass of the multiple systems from the main model (top) and the models M2 (bottom) at the end of the simulation.
The bottom plot shows all stars of the 5 models M2 together.
The real mass of a system is $m_{\rm rl} = m_1 + m_2$.
The observed mass of the apparent star is lower for systems with higher $q$.
The minimum of $m_\mathrm{obs}/m_\mathrm{rl}$ for binaries with both components on the main sequence is $\approx 0.55$.
All systems below this minimum are either of an order higher than two or include \acp{BH}, NSs, or other non-main-sequence stars, which are assigned a luminosity of $0 ~\mathrm{L_\odot}$.

The underestimation of the masses of stellar systems leads to an underestimate of the total masses of the \ac{SC} and its tails.
In all three components (\ac{SC}, \ac{T1}, and \ac{T2}) the fraction of higher-order multiples (more than two stars) is small.
In fact, in the main model, no higher-order multiple composed of only main-sequence stars is found in the last timestep.
While the number of binaries in the \ac{SC} and its tails differ, the qualitative distribution of binary masses is similar in all three components.

\subsection{Evolution of the binary fraction and the total photometric and dynamical mass}

\begin{figure*}
    \includegraphics[scale=1]{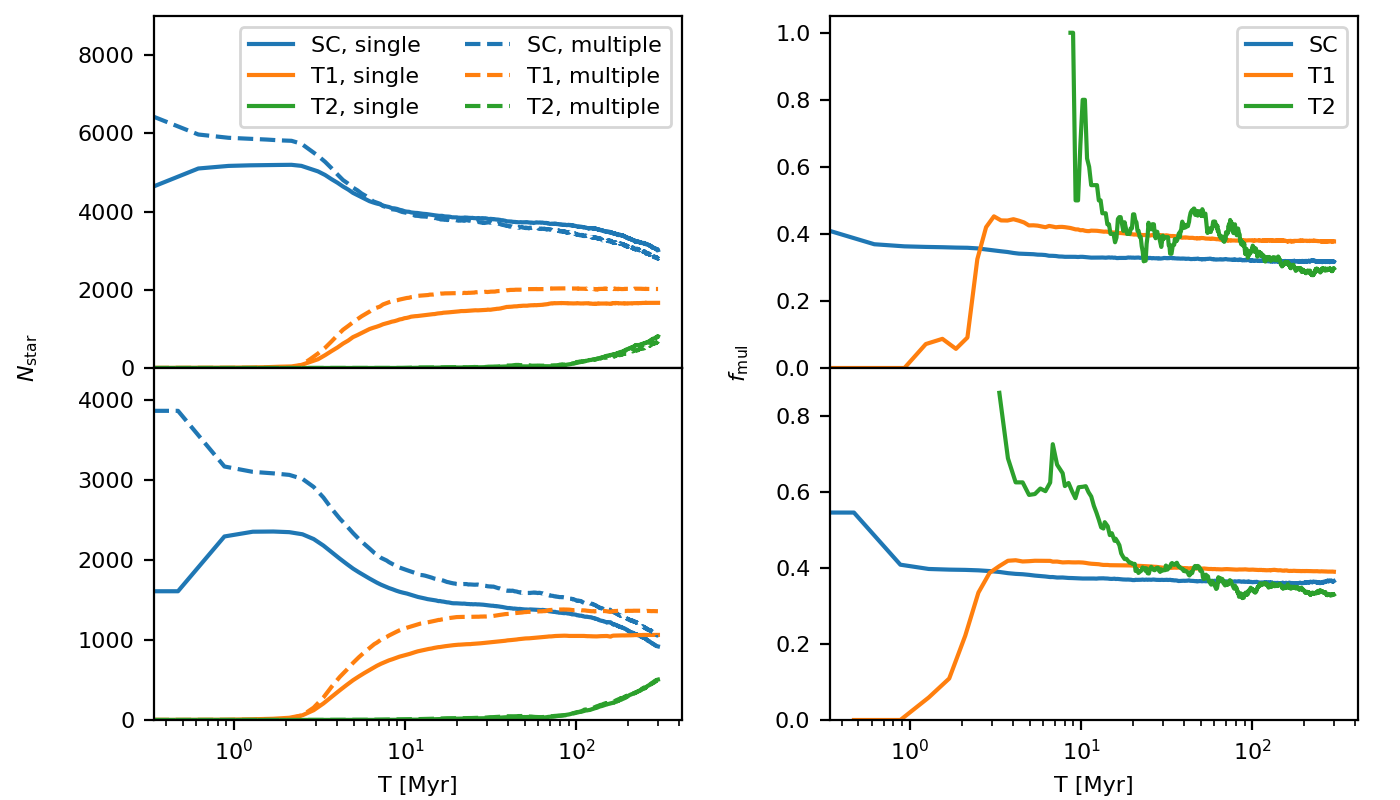}
    \caption{Numbers of single stars and stars in multiple systems for the SC, T1, and T2 (left).
    The fraction of  multiple systems over the total number of systems (counting single stars as systems), $f_{\rm mul}$ (Eq. \ref{eq_fmul}), computed from these numbers can be seen on the right.
    The top panels show the main model, while the bottom panels are for the five  M2 models (Table \ref{tab_modelParam}).}
    \label{fig_NumStars}
\end{figure*}

Figure \ref{fig_NumStars} shows the number of stars in single and multiple systems for the \ac{SC}, \ac{T1}, and \ac{T2} (left panel) and the fraction, $f_\mathrm{mul}$, of multiple systems over the total number of systems (right panel).
A system can be a single star or a multiple system.
We note that a vast majority of the multiple systems are binaries and only a few of them are triple and quadruple systems (see also Fig. \ref{fig_BinMasses}).
For correctness, they will still be referred to as multiple systems in this work.
The number of stars in multiple systems (initially close to 100\%; see also Fig. \ref{fig_NumStars_lin} for a linear plot that contains $T = 0$) decreases rapidly during the first megayear of cluster evolution, which is due to dynamical encounters in the dense environment \citep[see also][]{1995MNRAS.277.1491K,1995MNRAS.277.1507K}.
There are a few dynamical ejections from the \ac{SC} before gas expulsion starts at 0.6 Myr.

With gas expulsion \ac{T1} forms rapidly and inherits the fraction of stars in multiples from the young \ac{SC} models.
The low multiple fraction in \ac{T1} at the beginning of its formation is due to dynamical ejections as stars ejected dynamically before $t_\mathrm{ee}$ are counted towards \ac{T1}.
In the main model the higher density in the \ac{SC} leads to a decrease in the multiple fraction, while it remains constant at $f_{\rm mul} \approx 0.38$ in \ac{T1}.
\ac{T2} evaporates from the \ac{SC} more slowly and, therefore, at the end of the simulation, has almost the same fraction of multiples ($f_{\rm mul} \approx 0.31$) as the ageing \ac{SC} ($f_{\rm mul} \approx 0.32$).
During the formation of \ac{T2} the multiple fraction fluctuates due to low-number statistics.
Since the simulated \ac{SC} moves in the Galactic plane, the tails form in the Galactic plane as well.

Figure \ref{fig_NumStarsM3M4} shows the models M3 (top) and M4 (bottom).
Due to their lower densities and longer crossing times the binaries are dissolved more slowly and to a lesser extent the less massive a \ac{SC} becomes \citep{1995MNRAS.277.1491K,1995MNRAS.277.1507K}.
This leads to higher multiple fractions in \ac{T1} and the \ac{SC}.
It is also noticeable that the difference between \ac{T1} and the \ac{SC} is decreased.
This is because the time after \ac{T1} formed is not sufficient to reduce the multiple fraction in the \ac{SC} further.
It is also worth noting that in the models M3 and M4 the binary fraction in the \ac{SC} increases towards the end of the simulation.
This is due to the preferential evaporation of single stars.
The overall binary fraction in these models is decreasing up to the end.

However, the lower-mass models lose a larger fraction of their mass to \ac{T1}, so that at the end of the simulation the number of \ac{T1} stars is larger than the number of \ac{SC} stars.
This is consistent with their shorter relaxation times (see Table \ref{tab_results}).
This is caused by the less deep gravitational potential counteracting the tidal forces on the \ac{SC}.

\begin{figure}
    \includegraphics[scale=1]{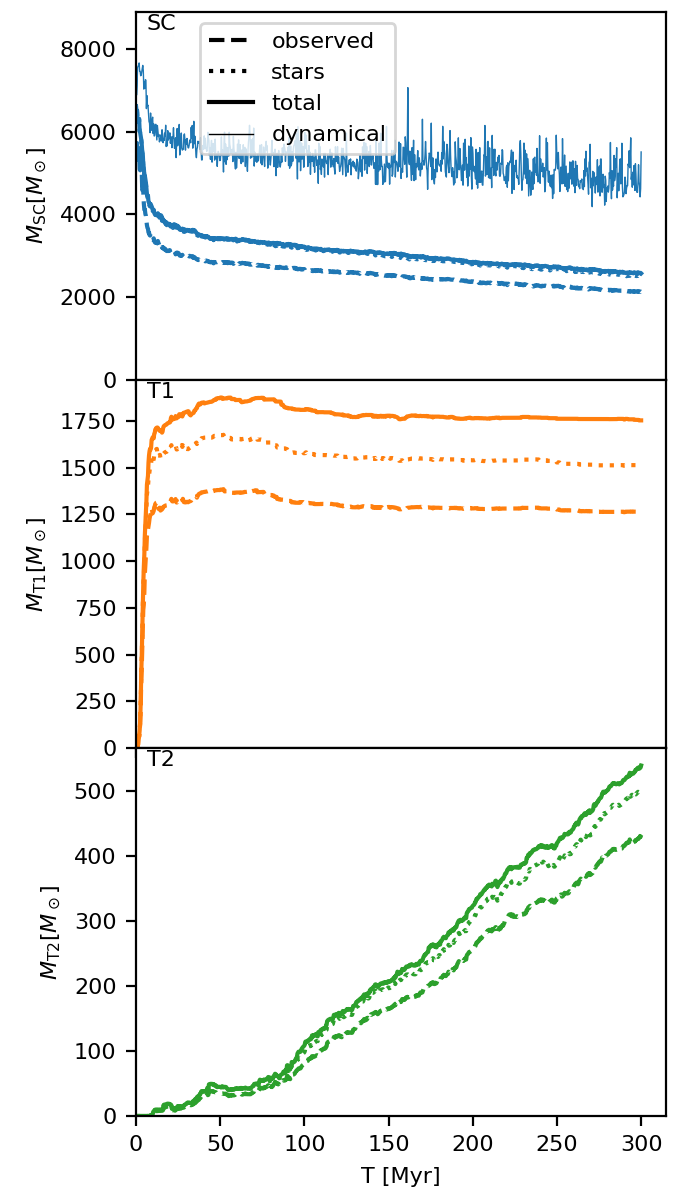}
    \caption{Development of the total mass, mass in main-sequence stars, and observed mass of the SC (top), T1 (middle), and T2 (bottom) of the main model over time.
    The dynamical mass estimate according to Eq. \ref{eq_dynmass} is only shown in the top panel for the \ac{SC}.}
    \label{fig_MassDevel}
\end{figure}

Figure \ref{fig_MassDevel} shows the development of the masses of the \ac{SC}, $M_{\rm SC}$, and the tidal tails, $M_{\rm T1}$ and $M_{\rm T2}$, of the main model.
The total mass (shown as the thick solid line) includes all stars, NSs, and \acp{BH}.
For the stars and observed masses only stars on the main sequence are taken into account.
This simplification is possible due to the short giant phases and low mass-fraction of high-mass stars in the \ac{SC} and its tails.
The mass white dwarfs would add leads to no noticeable change.
Therefore, white dwarfs are neglected as well.
The dashed curve takes into account that an observer would not only underestimate the \ac{SC} mass due to objects they cannot see, but also due to unresolved multiples.
As explained in Sect. \ref{sec_obsSysM}, the apparent mass of a multiple system is smaller than the sum of the masses of the companions, resulting in a lower `measured' total mass.
Therefore, this curve only includes stars on the main sequence and in the case of multiples the apparent mass of the star an observer would deduce from the luminosity of the unresolved system.
The exact values of the underestimate at the end of the simulations can be found in Table \ref{tab_results}.

The sum of the total masses of the \ac{SC} and its tails ($M_{\rm SC}^{\rm total} + M_{\rm T1}^{\rm total} + M_{\rm T2}^{\rm total}$) decreases over time due to mass loss through stellar evolution.
The \ac{SC} loses mass continuously through ejection, evaporation and, to a lesser extent, stellar evolution.
\ac{T1}, on the other hand, initially gains a lot of mass and then slowly loses mass due to stellar evolution only.
This is because gas expulsion only plays a role in the early stages of the evolution (see also Fig. \ref{fig_NumStars}), after which mass is lost due to stellar evolution.
\ac{T2} gains mass almost continuously, which means that evaporation of stars from the \ac{SC} adds mass faster than can be lost due to stellar evolution.

In addition to the underestimate of the \ac{SC}, \ac{T1}, and \ac{T2} masses due to the above-described bias from unresolved multiple systems, the amount by which the mass of a \ac{SC} would be overestimated using the velocity dispersion is computed.
As mentioned in the introduction, \citet{1976ApJ...204...73I} and \citet{1987degc.book.....S} found that the determined dynamical mass is proportional to the square of the velocity dispersion of the centres of mass of the stellar systems in the \ac{SC}, $\sigma_{\rm COM}$.
Following \citet{2020ApJ...896..152R} it is assumed that an observer would see the centre of light movement, rather than the centre of mass movement.
The proportionality factor between the cluster mass and the velocity dispersion depends on the assumed mass profile of the cluster.
To identify merely the error caused by the velocity dispersion, we therefore used
\begin{align}
    \label{eq_dynmass}
    M_{\rm SC}^{\rm dynamical} = M_{\rm SC}^{\rm total} \left( \frac{\sigma_{\rm COL}}{\sigma_{\rm COM}} \right)^2,
\end{align}
with $\sigma_{\rm COL}$ being the velocity dispersion of the centre of lights of the stellar systems.
For our calculations we use the velocity dispersions in z-direction ($\sigma_{\rm COL} = \sigma_{{\rm COL},z}$, $\sigma_{\rm COM} = \sigma_{{\rm COM},z}$).

Here it is important to determine which stars an observer would count as members of the \ac{SC}, as stars with too large radial velocities are often discarded as non-members.
To this end, the stars are sorted by their distance to the \ac{SC} centre.
For each star the velocity dispersion $\sigma_{{\rm COL},z}$ is computed from its own velocity and the velocities of its 20 neighbours in each direction.
The velocity dispersion is computed for each star individually to account for the velocity dispersion being larger in the centre of the \ac{SC}, than in the outskirts.
Systems whose velocities are larger than $3 \sigma_{{\rm COL},z}$ are excluded from the calculations.
This method is applied iteratively until no stars remain outside this variance.

As seen in Fig. \ref{fig_MassDevel}, the dynamical mass is significantly larger than the total mass of the SC.
For the main model, we measure an overestimate by a factor of 1.86 (Table \ref{tab_results}) at the end of the simulation.

At the same time, the mass of the \ac{SC} is underestimated by $3 ~\%$ due to the presence of unseen \acp{BH} and NSs and a further $15 ~\%$ due to unresolved binaries.
The effect caused by \acp{BH} and NSs is small since many of them get ejected from the \ac{SC} due to {SN} kicks.
Similarly, the mass in \ac{T1} (\ac{T2}) is underestimated by $14 ~\%$ ($6 ~\%$) due to not detecting \acp{BH} and NSs and a further $14 ~\%$ ($14 ~\%$) due to unresolved binaries.
Here, the effect of \acp{BH} and NSs is stronger, since the remnants kicked out from the \ac{SC} by SNe are added to its tails in our model.
However, since these remnants are kicked out with high velocities ($\approx 40 ~\rm km/s$), their phase space positions do not necessarily match those of the other tail members.
Since the most massive stars are short-lived and the majority of SNe explode early on, it is expected that dark objects play a more significant role in \ac{T1}'s mass underestimate, while in the \ac{SC} and \ac{T2} unresolved binaries cause the majority of the underestimate.

Figure \ref{fig_MassDevel_lowmass} shows a similar development for the Model M4.
Due to the higher binary fraction the dynamical overestimate of the \ac{SC} mass is higher than for the main model (17.52 compared to 1.86 for the main model; Table 2).
The underestimate of the \ac{SC} caused by \acp{BH}, NSs, and binary systems is similar to that of the main model at 22 \%.
We also note that the dynamical mass of the \ac{SC} increases at the end of the simulation together with the binary fraction.

\begin{figure}
    \includegraphics[scale=1]{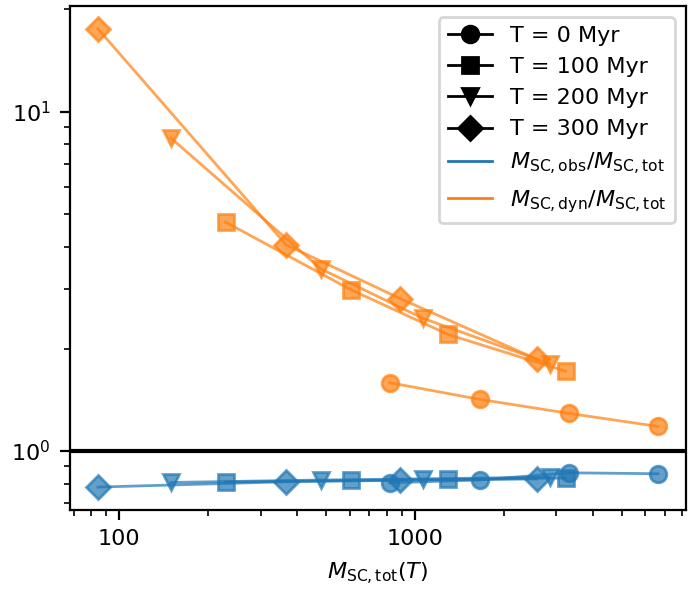}
    \caption{Over- and underestimate of the SC masses due to binary fractions and dark objects at different times depending on the cluster's current mass.
    The values are averaged over the last five timesteps, which span less than 1.3 Myr in total.}
    \label{fig_MassDevelAll}
\end{figure}

Figure \ref{fig_MassDevelAll} shows the under- and overestimates of the masses at different times for all models used in this work.
These values found for the overestimates using the dynamical mass estimate are smaller than the results found by \citet{2020ApJ...896..152R}, who for a cluster of a similar binary fraction as the ones shown in this work ($M_{\rm SC, tot} = 275 ~M_\odot, f_{\rm mul} = 0.421$) found that the velocity dispersion is overestimated by a factor of 6.7.
Due to the square in Eq. \ref{eq_dynmass} this would correspond to a mass overestimate by a factor of 44.8.
The main reason for that difference is that in the present work the stars with the largest velocities are removed as explained above, since this work assumes that an observer would probably not count these stars as cluster stars.
This leads to a smaller velocity dispersion in the present work.
A full list of the over- and underestimates of the \ac{SC} masses at the end of the simulations can be found in Table \ref{tab_results}.

Both the overestimate of the dynamical mass and the underestimate of the visible mass of the \acp{SC} become more pronounced over time.
It is visible that the dynamical mass is both mass and time dependent.
This is mainly due to the high initial cluster density leading to a large velocity dispersion.
If a \ac{SC} already has a large velocity dispersion, binaries are not increasing the perceived velocity dispersion by a large factor.
Low-density, low-mass clusters on the other hand have a low velocity dispersion, which means that binaries increase the perceived velocity dispersion and the dynamical mass by a large factor.

As mentioned in the introduction, observers often use the line-of-sight velocity dispersion to determine the dynamical mass instead of the photometric mass.
This work clearly demonstrates that not only is the dynamical mass overestimated due to binaries; the photometric mass is significantly underestimated.
This increases the apparent discrepancy between the cluster masses determined by different methods.
This is relevant for the Hyades star cluster, which is super-virial by a factor of about two \citep[][and references therein]{2022MNRAS.517.3613K}.
For the main model of this work, the dynamical and luminous mass differ initially by a factor of 1.6, which increases to 2.3 at the end of the simulation.
For the other models this factor is even larger.
However, many of the binaries in Hyades have been resolved \citep{1995AJ....110.1248G}, leaving a smaller difference than in our models that assume all binaries are unresolved.

\subsection{The velocity dispersion of Galactic field stars}
\label{sec_VeloDisp}

\begin{figure*}
    \includegraphics[scale=1]{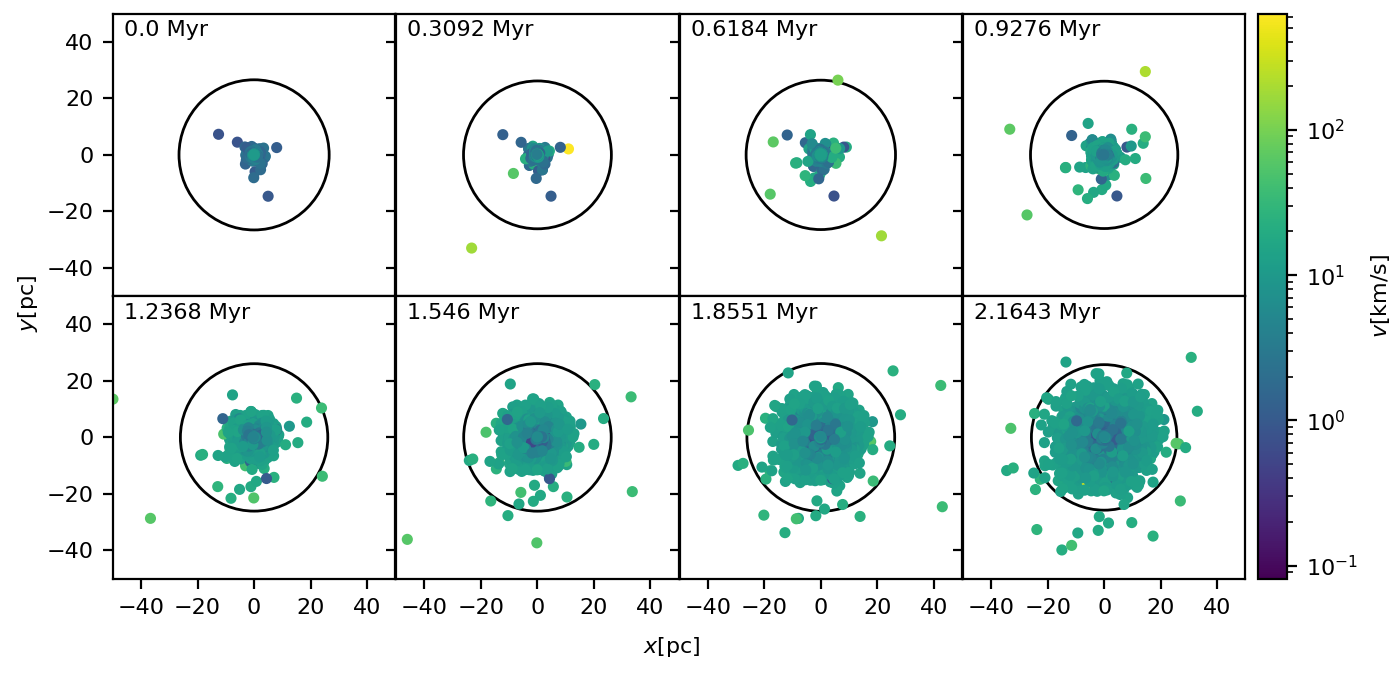}
    \caption{Snapshots of the SC (main model; Table \ref{tab_modelParam}) in the x-y plane (the Galactic centre being in negative x-direction) for the first eight timesteps to show the expansion due to gas expulsion.
    The tidal radius is shown as a black circle, and the stars are colour-coded according to the absolute value of their velocities relative to the SC's centre of mass.}
    \label{fig_Snapshots}
\end{figure*}

\begin{figure}
    \includegraphics[scale=1]{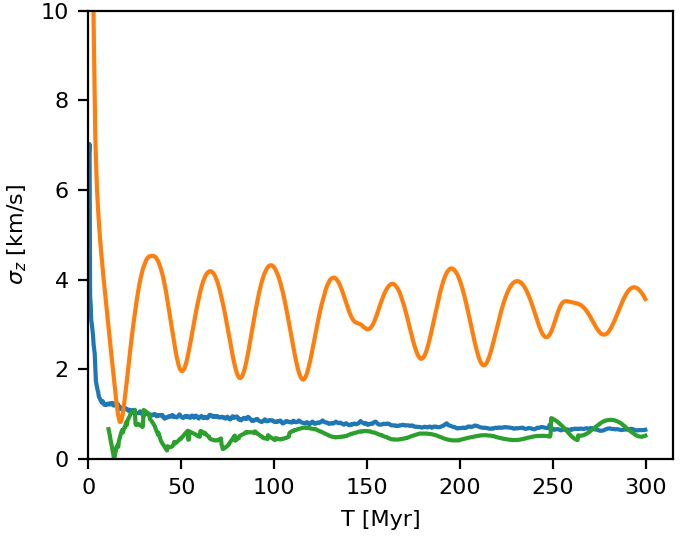}
    \caption{Velocity dispersion of stars in the z-direction for all stars between 0.3 and 0.7 $M_\odot$ in the SC (blue), T1 (orange), and T2 (green) for the main model.}
    \label{fig_VeloDispZ}
\end{figure}

By analysing star-count data, \citet{2023arXiv230107029L} deduced apparently systematic variations in the \ac{IMF} with the velocity dispersion in $z$-direction (perpendicular to the Galactic plane) and metallicity.
Such a variation in the \ac{IMF} has important astrophysical implications \citep{2013pss5.book..115K}.
Here we test if the origin of stars in star clusters may not partially or wholly account for this apparent variation.
This may result if the binary fraction and velocity dispersion conspire to lead to apparently systematic variations in the observationally deduced \ac{IMF}.

The velocity dispersion vertical to the Galactic plane is used as a proxy for the line-of-sight velocity dispersion.
This is done because the velocity ellipsoid in radial direction with respect to the Galactic centre is larger than in tangential direction.
However, the orientation of these axis in the non-rotating x-y plane changes over time.
Therefore, the velocity dispersion in x- and y-direction is expected to vary with time, as the population revolves around the Galaxy \citep[for a more detailed description of how the tails develop over time, see][]{2020A&A...640A..84D}.
In the case of multiples the velocity of the centre of mass is used for the calculations.
We note that the observer would observe the \ac{SC} primarily from within the Galactic disc and, therefore, would have to account for the positions of the \ac{SC} relative to the Sun.

Snapshots of the first eight timesteps of the main model are shown in Fig. \ref{fig_Snapshots}.
The tidal radius, calculated according to Eq. 1 from \cite{2003MNRAS.340..227B} is shown as a black circle.
Only one star escapes the \ac{SC} at 0.3 Myr, which is clearly dynamically ejected.
However, in the same timestep we also see a star within the cluster with a velocity exceeding $100 ~\mathrm{km/s}$.
This is typical for an ejected star about to leave.
After 0.6 Myr it is difficult to distinguish between stars that left the \ac{SC} primarily due to gas expulsion and those leaving primarily due to dynamical effects.
It is clear, however, that only a handful of stars leave the \ac{SC} before gas expulsion.
These stars leaving the \ac{SC} propagate with a velocity of a few tens to hundreds of km/s.

The velocity dispersion for stellar systems in the \ac{SC} and its tails is shown in Fig. \ref{fig_VeloDispZ}.
The high peak in the velocity dispersion at the very beginning for both the \ac{SC} and \ac{T1} is caused by stars being ejected due to stellar-dynamical encounters.
As gas expulsion starts the stars lost due to gas expulsion quickly form the majority of the \ac{T1} stars.
These stars oscillate around the Galactic plane.
Since the \ac{T1} stars are all expelled during a very short time interval through gas expulsion, this translates to oscillations of the velocity dispersion for \ac{T1} around about 3 km/s.
They can be distinguished easily from the stars in the \ac{SC} and \ac{T2} due to their larger velocity dispersion.

The \ac{T1} stars complete about 9.9 vertical oscillations over the 300 Myr of simulation.
This translates to a period of about 32 Myr, which is comparable to the vertical period of the Sun around the Galactic plane \citep{1985Natur.316..706B}.
\Ac{T1} ends up far more dispersed in both the Galactic plane and the orthogonal direction, as can be seen from the velocity dispersions.
Therefore, \ac{T2} would be expected to be much easier to observe than \ac{T1}.

As can be seen in Table \ref{tab_results} differences between the velocity dispersions of different models are small compared to the differences between \ac{T1} and \ac{T2}.
More massive \acp{SC} produce tidal tails with fewer binaries and a larger velocity dispersion.
Only looking at this result, it might seem as if field populations with a higher velocity dispersion would automatically be expected to host fewer binaries.
However, we have also seen that \ac{T1} hosts a larger fraction of binaries and has a larger velocity dispersion than \ac{T2}.
For the models investigated here, the difference between different tails is larger than the difference produced by different \ac{SC} masses.
To get a complete quantification for a correlation of binary fraction and velocity dispersion in the Galactic field a complete simulation of \acp{SC} following the MF of embedded star clusters is needed.
However, it should be noted that \citet{2023arXiv230107029L} do find a higher binary fraction in dynamically hot stellar populations than in cold populations observed using LAMOST.
They distinguish the two groups of stars by their action in the z-direction, $J_z$, and call those with $J_z > 20 ~\rm kpc \; km / s$ dynamically hot.
Stars with line-of-sight velocities exceeding ${25 ~\rm km / s}$ are assumed to be in binaries by them.

\begin{figure}
    \includegraphics[scale=0.95]{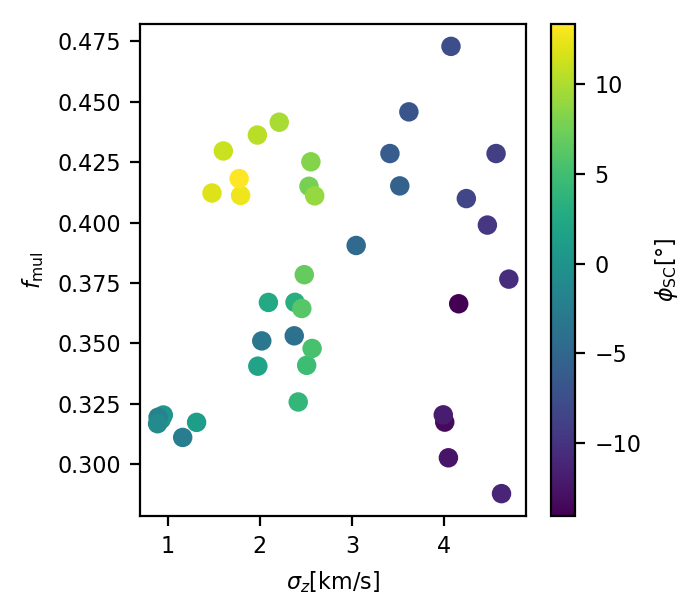}
    \caption{Fraction of stars in multiple systems over the velocity dispersion of stars in the z-direction at the end of the simulation (T = 300 Myr) for the main model. The points are colour-coded according to the angular distance in degrees of the cylinder from the SC with respect to the Galactic centre.
    Positive values of $\phi_{\rm SC}$ indicate the stellar populations in the leading T1 and T2 tails, while $\phi_{\rm SC} < 0$ indicate the trailing tail ones.}
    \label{fig_VeloDispFbin}
\end{figure}

\begin{figure}
    \includegraphics[scale=0.95]{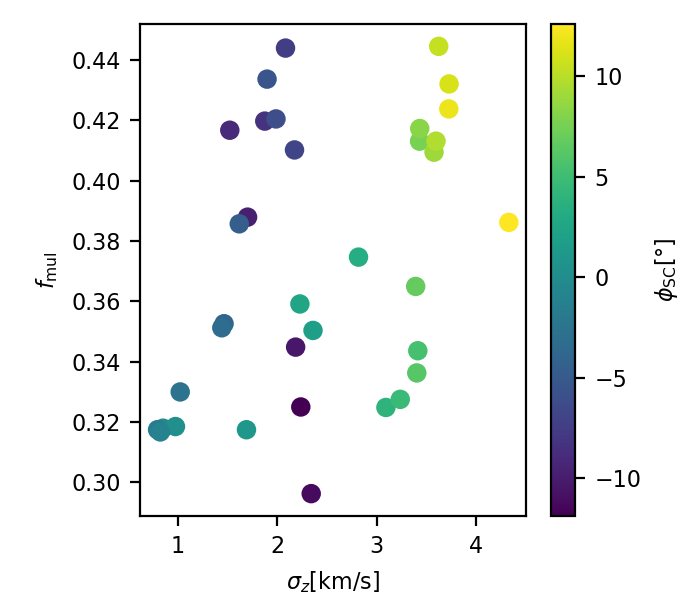}
    \caption{Same as Fig. \ref{fig_VeloDispFbin}, but at $T = 283 ~\rm Myr$.}
    \label{fig_VeloDispFbin283}
\end{figure}

Combining the previous results -- that the binary fraction in \ac{T1} is larger than in \ac{T2} and the \ac{SC} -- with the fact that the stars in \ac{T1} have the largest $\sigma_\mathrm{z}$, it follows that the binary fraction in groups of stars belonging to the same tail is expected to be larger if $\sigma_\mathrm{z}$ is large, and vice versa.
In Fig. \ref{fig_VeloDispFbin} the fraction of multiple systems, $f_\mathrm{mul}$, is shown depending on the velocity dispersion.
To obtain this plot several cylindrical volumes with a radius of 250 pc and infinite length, oriented orthogonal to the Galactic plane, are inserted into the tidal tails at different angular distance from the \ac{SC} as seen from the Galactic centre.
These cylinders overlap.
The reason for using cylinders is that the velocity dispersion in x- and y-direction (in the non-rotating frame) is affected by the Galactic rotation, while in z-direction it is independent of it.
Within each of those cylinders the fraction of multiple stars, $f_\mathrm{mul}$, and $\sigma_{z}$ is measured.
For each timestep, 100 cylinders are used; however, they are ignored if they contained fewer than 100 systems.
From the colour coding it is visible that both $\sigma_\mathrm{z}$ and $f_\mathrm{mul}$ are increasing with increasing distance to the \ac{SC}.
However, the spread also increases significantly, so that it would become difficult for an observer to detect a correlation between $\sigma_\mathrm{z}$ and $f_\mathrm{mul}$ at large distances to the \ac{SC}.
Especially in the trailing tail, the number of binaries seems to decrease again at large distances (10°) to the SC.
However, this decrease corresponds to about 10 pairs per cylinder.
Because of their size, the cylinders overlap, so that the decrease in the binary fraction is likely caused by a local under-density of binaries in this particular model snapshot.

The velocity dispersion in this snapshot (shown in Fig. \ref{fig_VeloDispFbin}) is larger in the trailing tail, than it is in the leading tail.
However, as already shown in Fig. \ref{fig_VeloDispZ} the velocity dispersion in \ac{T1} fluctuates over time. In Fig. \ref{fig_VeloDispFbin283}, the same plot as Fig. \ref{fig_VeloDispFbin} is shown for $T = 283 ~\rm Myr$.
Here the leading tail has a larger velocity dispersion than the trailing one.
The phase difference is caused by the leading tail initially moving towards the Galactic centre and the trailing tail moving away from it.

Overall these results show that there is a correlation between binary fraction and $\sigma_\mathrm{z}$.
However, a stellar-dynamical population model of the Galactic field population, consisting of many accumulated populations, would need to be computed to estimate systematic biases in deriving the stellar \ac{IMF} from star counts.

\subsection{The effect of the binary fraction on the stellar \ac{MF}}
\label{sec_EffectMF}

\begin{figure}
    \includegraphics[scale=1]{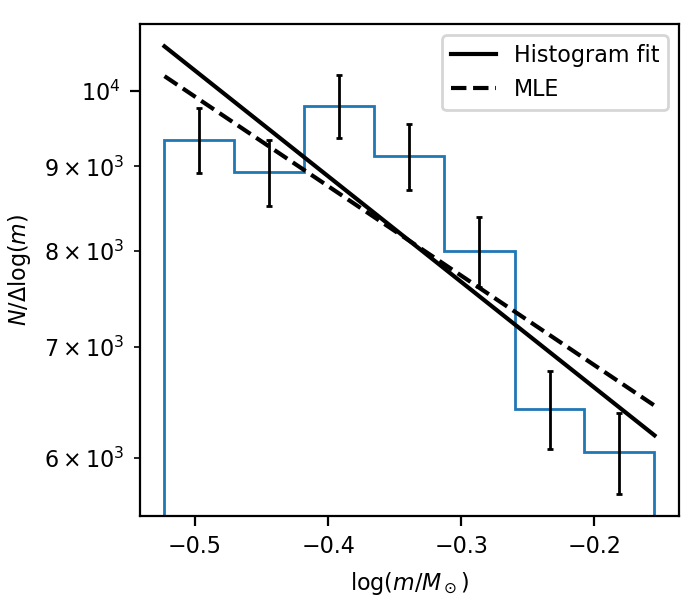}
    \caption{Bins used to fit the MF between $0.3$ and $0.7 ~M_\odot$ to the stars of the SC in the initial timestep ($T = 0$) of the main model.
    The fitted MF is shown with a solid line, and the MF determined using the MLE (with k fitted to best fit the bins) with a dashed line.
    The error bars show the Poisson uncertainties.}
    \label{fig_ExMFfit}
\end{figure}

The stellar \ac{MF} is fitted to the stars in the generic mass range between 0.3 and 0.7 $M_\odot$.
This is the typical mass range used by other researchers \citep[see Table 2 in][]{2024ApJ...969...95Y,2000ApJ...534..870P,2018MNRAS.478.1520B,2022arXiv220503323L}.
The low-mass end is limited by completeness, while the high-mass end serves to remove stellar evolution issues.
For the entire simulated period of 300 Myr, these stars never leave the main sequence.
Therefore, no post-main-sequence objects need to be taken into account and the \ac{MLR} from \citet{1993MNRAS.262..545K} can be used.
In this work, two main methods for deriving the power-law index, $\alpha,$ were employed:

The first method is histogram fitting, which was mentioned by \citet{2009MNRAS.395..931M}.
It is visualised in Fig. \ref{fig_ExMFfit}.
The stars are first sorted into logarithmic mass bins and their number is divided by the bin width, $w = \log_{10} (m_{j + 0.5}) - \log_{10} (m_{j - 0.5})$ ($m$ is in units of $M_\odot$).
$j$ is the number of the bin and $w$ is equal for all bins.
The mass \ac{MF} $dN = \xi(m) dm = km^{-\alpha} dm$ is a power-law, which becomes linear in log-log.
Therefore, we translated the \ac{MF} into the double logarithmic frame and obtain for each bar in the histogram at the position $\log_{10}(m)$ of height $N_j$:
\begin{align}
        \xi( \log_{10}(m) ) &= k 10^{(1 - \alpha) \log_{10} (m)} \ln( 10 ),\\
        N_j &= \int\limits_{\log_{10}(m_{j-0.5})}^{\log_{10}(m_{j+0.5})} \xi( \log_{10}(m) ) d\log_{10}(m)\\
        &= \frac{k}{1-\alpha} \left( 10^{(1-\alpha)\log(m_{j+0.5})} - 10^{(1-\alpha)\log(m_{j-0.5})} \right),\\
        &= \frac{k}{1-\alpha} 10^{(1 - \alpha) \log_{10}(m_j)} \left[ 10^{(1 - \alpha) w/2} - 10^{-(1 - \alpha) w/2} \right],\\
    \label{eq_logNlogm}
    \log_{10} \left( N_j \right) &= (1 - \alpha) \log_{10}(m_j) + const.
\end{align}
The function in Eq. \ref{eq_logNlogm} is then fitted to the data to obtain the \ac{MF}.

The second method is the \ac{MLE} described in Appendix A of \citet{2013MNRAS.434.3236K}.
The \ac{MLE} is based on the work of \citet{2009SIAMR..51..661C,2009MNRAS.395..931M} who developed it for cases with only a lower boundary.
Their main idea is to formulate the likelihood that a certain distribution of data points results from a \ac{MF} with a given $\alpha$ and then maximising the function for $\alpha$.
As a first step both the fitting method and the \ac{MLE} are tested on an optimally sampled canonical \ac{IMF} \citep{2013pss5.book..115K,2017A&A...607A.126Y} between $m = 0.3$ and $0.7 ~M_\odot$.
This gives $\alpha = 1.65 \pm 0.03$ for the fitting approach above and $\alpha = 1.62 \pm 0.03$ using the \ac{MLE}.
It is important to note here that there is no `correct' value for the power-law slope here, since the true slope is a two-part power-law instead of the one-part power-law that is fitted here.
The lower value from the \ac{MLE} stems from the \ac{MLE} putting more emphasis on matching the low-mass portion of the \ac{MF} than fitting a histogram does.
If the low-mass portion of the \ac{MF} would have a larger $\alpha$ than the high-mass one we find that $\alpha$ determined with the \ac{MLE} to be larger than the one deduced from fitting a histogram.

\begin{figure*}
    \includegraphics[scale=1]{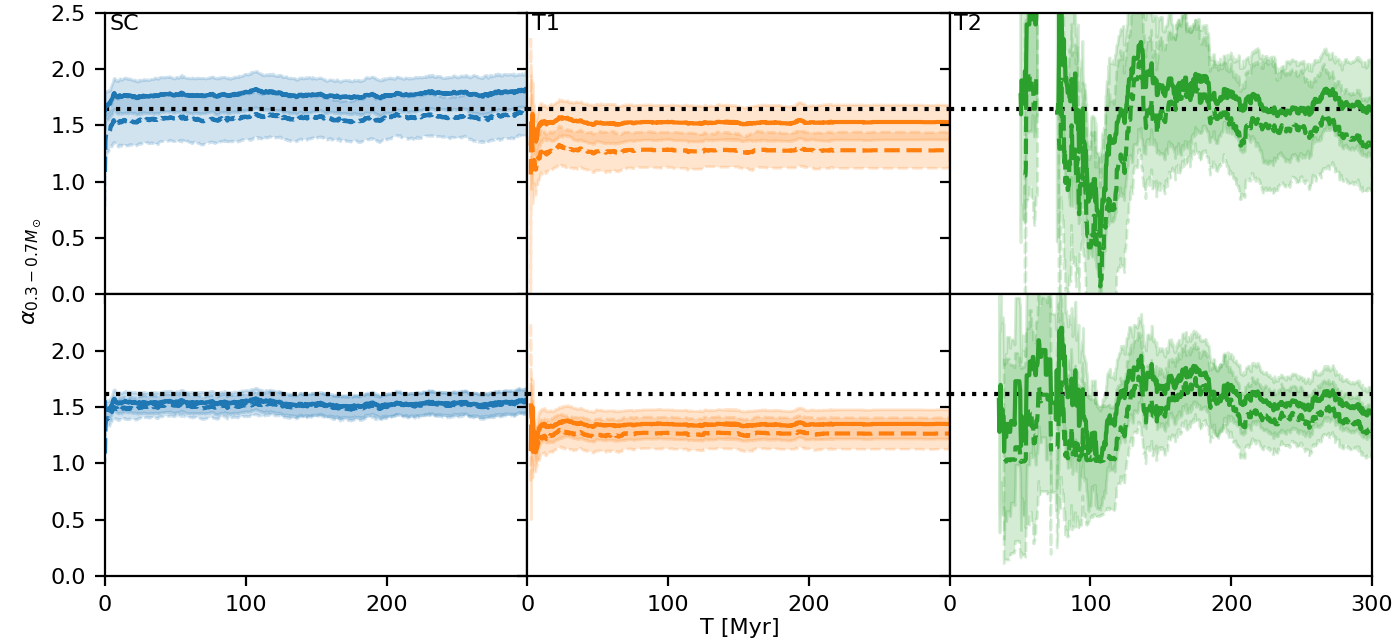}
    \caption{MF slope obtained using the fitting approach (Eq. \ref{eq_logNlogm}, top) and the MLE (bottom) between 0.3 and 0.7 $M_\odot$ over time for the main model.
    Both the MFs as seen by an observer who cannot resolve the binaries (dashed) and the actual real MFs (solid) of the resolved SC, T1, and T2 are shown.
    The coloured areas show the 1$\sigma$ confidence span.
    The canonical value of 1.65 is shown as a dotted black line.}
    \label{fig_Time-Alpha2}
\end{figure*}

Figure \ref{fig_Time-Alpha2} shows how the power-law index of the \ac{MF} for stars between 0.3 and 0.7 $M_\odot$, $\alpha_{\mathrm{ 0.3-0.7 M_\odot}}$, develops over time.
In all 3 groups (\ac{SC}, \ac{T1} and \ac{T2}) the observed \ac{MF} is flatter than the real one that would be measured if all binaries were resolved by about $\Delta\alpha = 0.2$.
For the main model the real \ac{MF} for stars in the \ac{SC} is steeper than the one for \ac{T1} ($\alpha_{\mathrm{ 0.3-0.7 M_\odot}}$ = 1.8 v.s. 1.5).
For the \ac{SC} and \ac{T1} the real \ac{MF} stays constant for the majority of the simulation.
Only in the very beginning, when \ac{T1} starts forming, are variations visible.
However, the \ac{MF} of the \ac{SC} is not always steeper than that of \ac{T1}.
Looking at the lower-mass models, we find no systematic tendency in either direction.
The observed variations can be attributed to the \ac{MF} not being optimally, but rather stochastically, sampled in the models used here and the models being non-mass-segregated \citep{2013pss5.book..115K}.
Compared to the \ac{MF} of the \ac{SC} and \ac{T1}, the \ac{MF} of \ac{T2} varies significantly.
This is due to the slow evaporation process that forms \ac{T2}.
As more stars join \ac{T2}, the variations decrease and the \ac{MF} of \ac{T2} becomes comparable to the other two.

The lower panels of Fig. \ref{fig_Time-Alpha2} show the $\alpha$-values for the same data as the upper panels, but obtained using the \ac{MLE}.
In all cases the real and observed values for $\alpha$ are closer together.
It is also visible that $\alpha$ of the \ac{SC} and \ac{T1} remains below the value obtained for an optimally sampled \ac{IMF}, despite this not being the case if histograms are used to determine $\alpha$.
This is a result of the clusters being stochastically rather than optimally sampled.
These differences in the direction in which the $\alpha$ value of a stochastically sampled \ac{MF} deviates from the optimally sampled one is the result of the \ac{MLE} putting more emphasis on the low-mass portion of the \ac{MF}.

\begin{figure*}
    \includegraphics[scale=1]{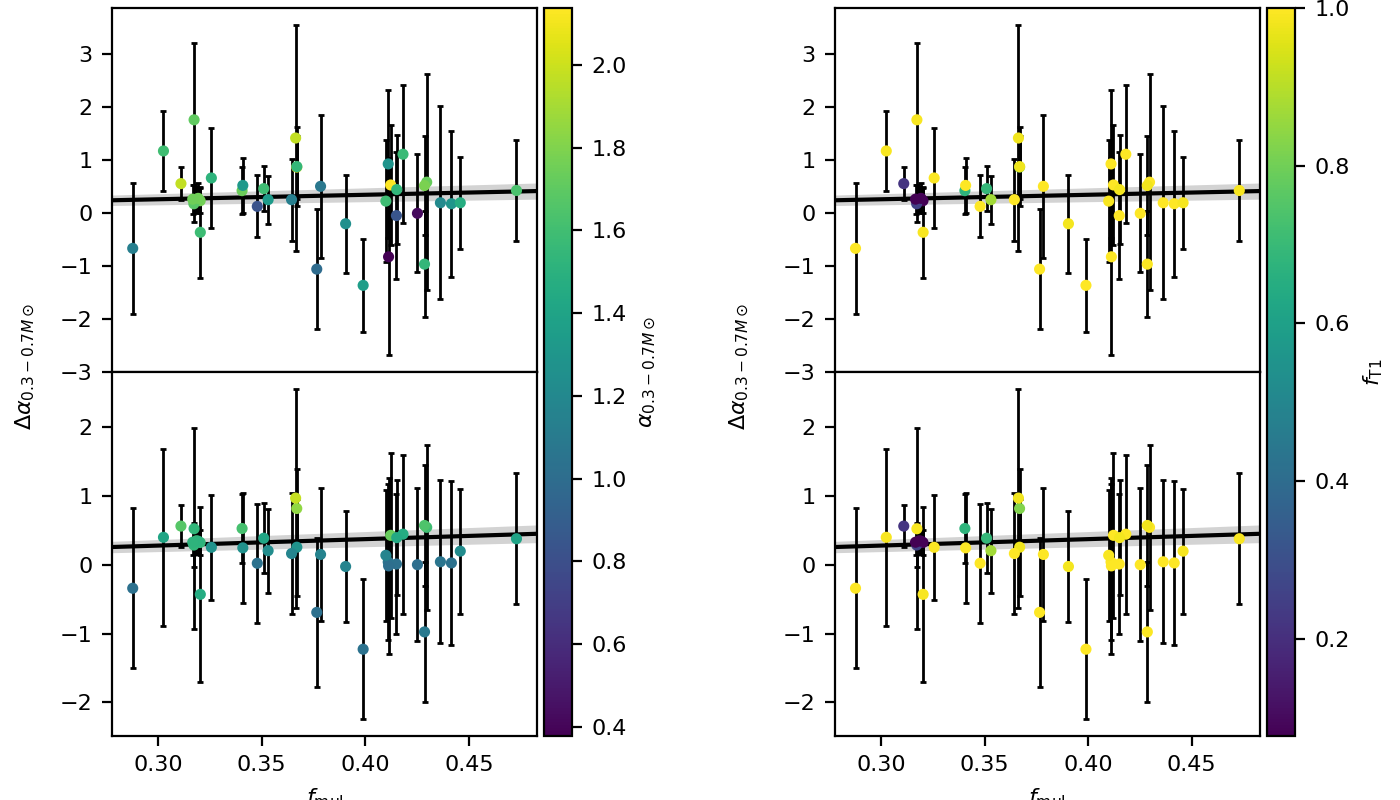}
    \caption{Difference between the observed and real $\alpha$ between 0.3 and 0.7 $M_\odot$ depending on the multiplicity for the last timestep of the main model (top) and the low-mass models (bottom).
    The data is colour-coded according to the real $\alpha$ (left) and the portion of T1 stars (right).
    Populations with a larger fraction of unresolved multiple systems thus have a larger $\Delta\alpha$ and therefore a smaller observed $\alpha$, i.e. the observer deduces a flatter MF.
    The grey area shows the uncertainty in the slope.
    Note that for the five low-mass models, all data points are plotted and fitted together.}
    \label{fig_fMul-alpha}
\end{figure*}

Figure \ref{fig_fMul-alpha} shows how the fraction of stars in multiples affects the difference between the real $\alpha$ of the population and the $\alpha$ an observer who cannot resolve the binaries measures, $\Delta\alpha = \alpha_{\mathrm{rl}} - \alpha_{\mathrm{obs}}$ for the cylinders introduced in Sect. \ref{sec_VeloDisp}.
The values of $\alpha$ used here were obtained via fitting a histogram (top) or using the \ac{MLE} (bottom).
Using a weighted least square fit, a linear function, $\Delta\alpha_{0.3-0.7 M_\odot} = b f_{\rm mul}$, was fitted.
$\chi^2$ for a linear function, $y = f(x) = bx$ is
\begin{align}
    \chi^2 = \sum_{i=1}^{N} \frac{(y_i - b x_i)^2}{\sigma_i^2},
\end{align}
with the fitting parameter, $b$, and the error for $y_i$, $\sigma_i$.
From this follows
\begin{align}
    b& = \frac{\sum\limits_{i=1}^{N} \frac{x_i y_i}{\sigma_i^2}}{\sum\limits_{i=1}^{N} \frac{x_i^2}{\sigma_i^2}} & \text{and} &
    &\Delta b &= \frac{1}{\sqrt{\sum\limits_{i=1}^{N} \left( \frac{x_i}{\sigma_i} \right)^2}}.
    \label{eq_linfit}
\end{align}
Using $f_{\rm mul}$ as $x$ and $\Delta\alpha_{0.3-0.7 M_\odot}$ as $y$, this yields $b = 0.8 \pm 0.3$ for the main model.
The values for all models are listed in Table \ref{tab_results}.
The positive correlation matches the results by \citet{2023arXiv230107029L} qualitatively, but not quantitatively.
This is due to largely different methods being used in their work, compared to the present one.
In addition to using a mass range between $0.1$ and $3.0 M_\odot$, compared to the $0.3$ to $0.7 M_\odot$ used in the present work, \citet{2023arXiv230107029L} neglected the luminosity contributed by the secondary instead of adding the luminosities.

The colour coding in the left panel of Fig. \ref{fig_fMul-alpha} shows that the multiplicity and $\Delta\alpha_{0.3-0.7 M_\odot}$ are larger for cylinders with a larger real $\alpha$.
This seems counter-intuitive, since \ac{T1} has a flatter observed \ac{MF} and higher multiplicity than \ac{T2} (see Figs. \ref{fig_NumStars} and \ref{fig_Time-Alpha2}).
However, looking at the right panel it becomes clear that the internal spread within \ac{T1} dominates over the differences between \ac{T1} and \ac{T2}.
Due to the stars in \ac{T1} being spread out more than the stars in \ac{T2}, the number of stars per cylinder is much lower.
This is reflected in the large error bars of the cylinders containing mainly \ac{T1} stars.

\section{Conclusions}
\label{sec_concl}

In the present work, realistic N-body simulations of \acp{SC} with masses between $800 M_\odot$ and $6400 M_\odot$ carried out with \textsc{nbody6} were used to determine the effect of multiple star systems on the measured masses and \acp{MF}.
These masses are near the upper limit of masses of nearby \acp{SC} \citep{2022MNRAS.516.5637E}.
The apparent mass a multiple system would have was computed by adding the luminosities produced by the companion stars and using the \ac{MLR} from \citet[see also \citealt{2002AJ....124.2721R}]{1993MNRAS.262..545K}.

The main results are as follows:
\begin{enumerate}[wide, label=(\arabic*), labelwidth=!, labelindent=\parindent]
    \item Stars in \ac{T1} have a higher fraction of multiples than those in \ac{T2} and the \ac{SC}.
    This is due to the \ac{T1} stars being expelled early, before the multiple fraction in the \ac{SC} drops to its final value.
    \item Due to the presence of remnants (NSs and \acp{BH}) and unresolved multiples, the photometric mass underestimates the total mass of the \ac{SC} by nearly one-fifth.
    Binaries are the main contributor to this effect.
    In the main model, they cause a 14 - 15 \% underestimate of the total photometric mass of the \ac{SC} and its tails. \Acp{BH} and NSs have only a small effect in the \ac{SC}, where they cause a 3 \% underestimate, but a very strong effect in \ac{T1}, the total photometric mass of which is underestimated by an additional 14 \% due to dark objects.
    The dynamical mass of the \ac{SC} in the main model is a factor of 1.6 higher than the luminous mass at the beginning of the simulation,  and a factor of 2.3 higher at its end.
    This is reminiscent of the Hyades-mass problem discussed in \citet{2011A&A...531A..92R} and \citet{2022MNRAS.517.3613K}.
    \item The \ac{MF} in the stellar mass range 0.3 - 0.7 $M_\odot$ appears more top-heavy to an observer who cannot resolve multiple systems.
    In both tails and the \ac{SC}, the difference $\Delta\alpha_{0.3-0.7 M_\odot}$ is $0.2 \pm 0.3$.
    While this is within 1$\sigma$, the fact that this difference is present at all timesteps, in both the \ac{SC} and its tails and in all simulations, suggests a systematic effect.
    \item While there is a positive correlation between $f_{\rm mul}$, $\sigma_\mathrm{z}$, and $\Delta\alpha_{0.3-0.7 M_\odot}$ within a single model, when looking at different models one can find a positive correlation between the initial \ac{SC} mass and the velocity dispersion in the tidal tails.
    However, the correlation between the \ac{SC} mass and binary fraction is negative.
\end{enumerate}

In this work it is assumed that all binaries are unresolved; however, in reality some binaries might be resolved and some not.
Therefore, future studies should take the distance of the \ac{SC} to the observer into account when studying these effects.

The field stars we observe in the Galaxy consist of a mixture of tidal tails of \acp{SC} of many different initial masses.
To recreate the stellar population actually observed in the field, a full population synthesis that includes a range of models with different \ac{SC} masses would be necessary \citep[see e.g.][]{2002MNRAS.330..707K,2011MNRAS.417.1702M,2022MNRAS.510..413D}.
This would have to be done by including a range of \acp{IMF} that reflect their dependence on the metallicity and density of the gas cloud the \ac{SC} is born from, as was found by comparing the results of N-body models to observed \ac{SC} properties, such as their masses, half-mass radii, and binary fractions \citep{2012MNRAS.422.2246M}, and by comparing the chemical evolution of galaxies computed for different \ac{IMF}s to the metallicities observed in dwarf galaxies \citep{2009ApJ...706..599L,2011MNRAS.415.1647G,2020A&A...637A..68Y}.
The two methods lead to similar results for the upper end of the \ac{IMF}.
It is also important to mention that \citet{2023MNRAS.521.3991B} measured the upper end of the \ac{IMF} of 14 Small and Large Magellanic Cloud clusters directly.
While the masses and metallicities of these clusters should theoretically lead to a flattened \ac{IMF}, their errors are too large to detect any systematic variation.
Similarly, \citet{2023MNRAS.522.5320D} investigated 37 Galactic globular clusters without finding any evidence for a variation in the \ac{IMF} after fitting multi-mass models to them.
However, they only cite statistical errors and, in contrast to the analysis by \citet{2012MNRAS.422.2246M}, \citet{2023MNRAS.522.5320D} neglect very early stellar dynamical processes during and before gas expulsion, which makes it difficult to assess the actual errors of the found \ac{MF} slopes.
Gas expulsion itself is not taken into account in these models.

The work presented here thus points towards the required deep analysis needing to be performed when studying the field-star population in order to infer how the stellar \ac{IMF} varies with the physical conditions of star formation.

\begin{acknowledgements}
We thank an anonymous referee for many helpful suggestions.
The authors acknowledge support from the Grant Agency of the Czech Republic under grant number 20-21855S and through the DAAD-Eastern-Europe Exchange grant at Bonn University.
HW thanks Roberto Capuzzo Dolcetta for valuable discussions.
\end{acknowledgements}



\bibliographystyle{aa}
\bibliography{BinSC} 

\begin{appendix}

\onecolumn
\section{Supplementary figures}

\begin{figure*}[h!]
    \includegraphics[scale=1]{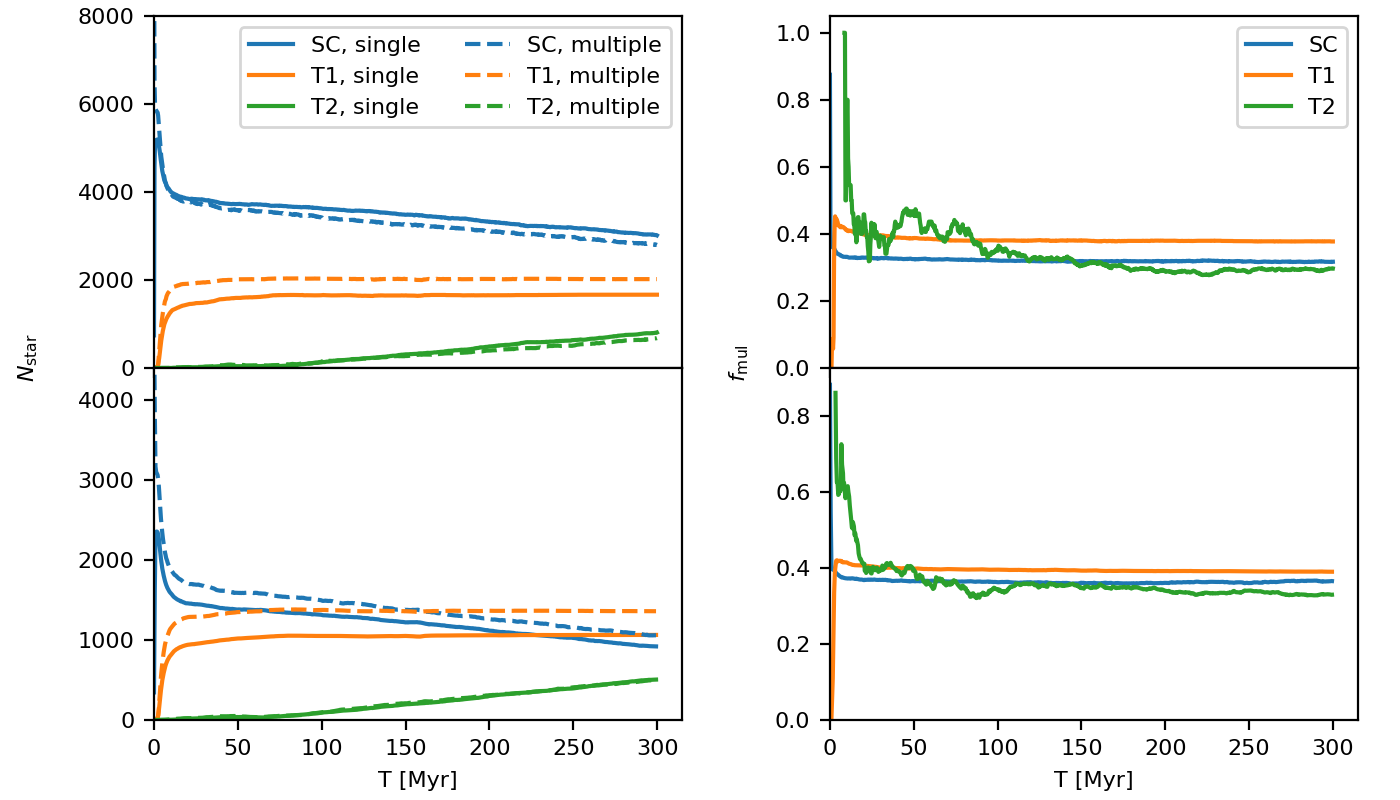}
    \caption{Same as Fig. \ref{fig_NumStars}, but with a linear timescale.}
    \label{fig_NumStars_lin}
\end{figure*}
\begin{figure*}[h!]
    \includegraphics[scale=1]{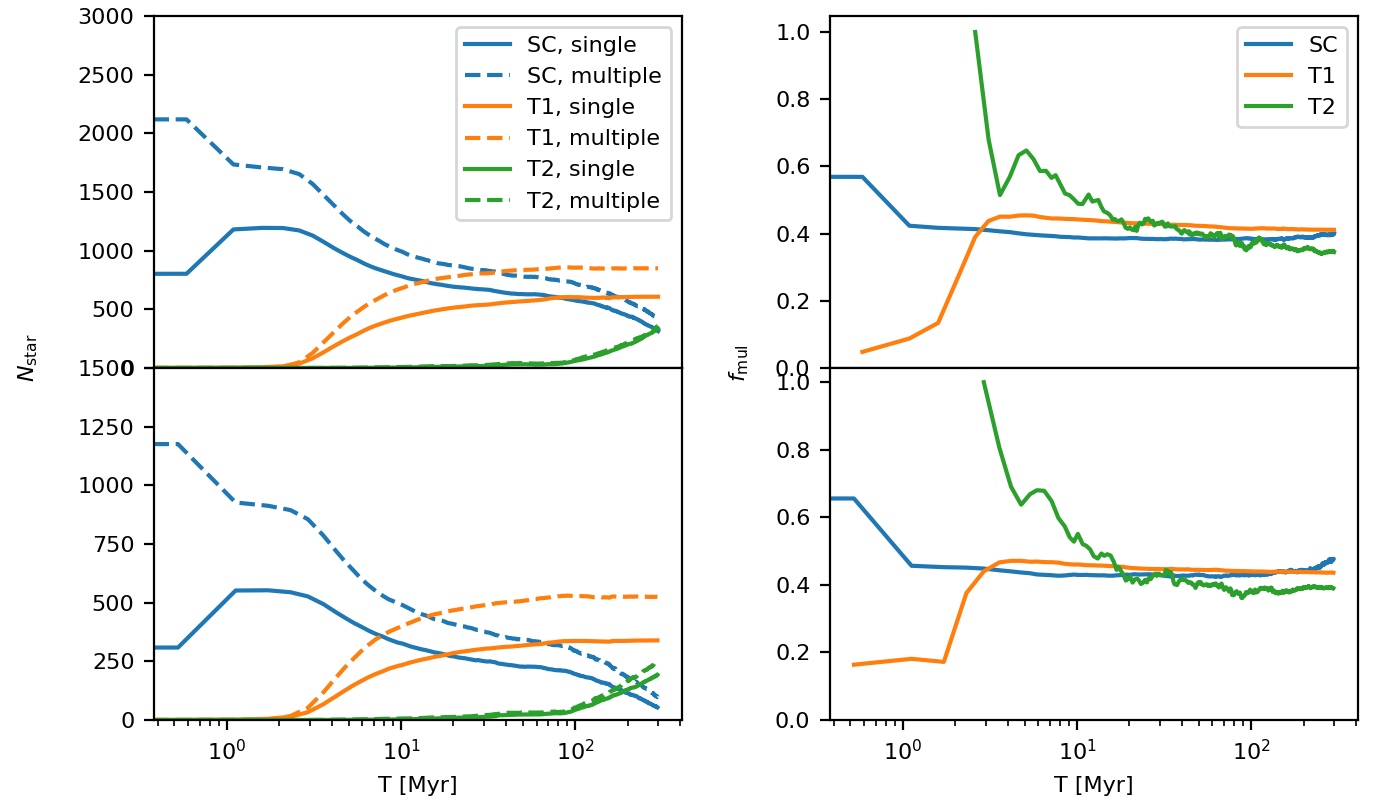}
    \caption{Same as Fig. \ref{fig_NumStars}, but for the models M3 (top) and M4 (bottom).}
    \label{fig_NumStarsM3M4}
\end{figure*}
\twocolumn

\begin{figure}
    \includegraphics[scale=1]{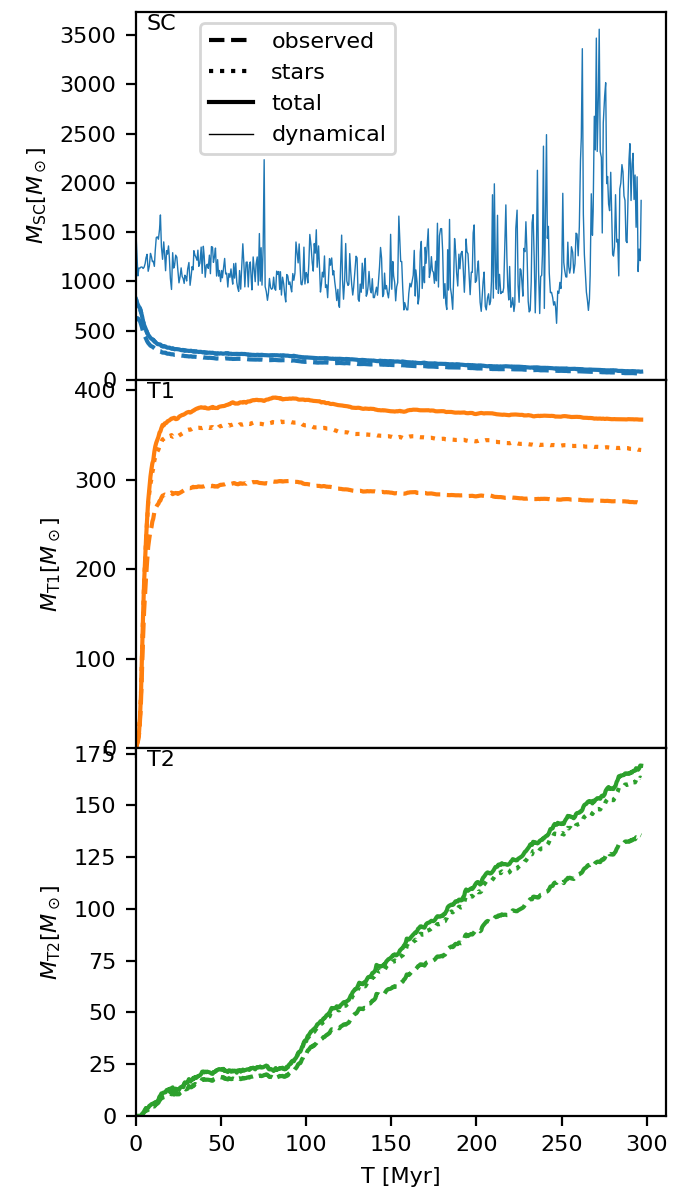}
    \caption{Same as Fig. \ref{fig_MassDevel}, but the average for the 16 M4 models  (Table \ref{tab_modelParam}).}
    \label{fig_MassDevel_lowmass}
\end{figure}

\end{appendix}

\end{document}